\documentstyle[preprint,aps,epsf,epsfig]{revtex}
\tightenlines
\newcommand{\bwf}[1]{\Psi^{#1}}
\begin{document}
\draft
 
\title{Dynamical Test of Constituent Quark Models with $\pi N$ 
Reactions}
\author{T. Yoshimoto$^1$, T. Sato$^1$, M. Arima$^2$, and T.-S. H. Lee$^3$}
\address{
$^1$ Department of Physics, Osaka University, Toyonaka, Osaka, 560-0043, Japan
\linebreak
$^2$ Department of Physics, Osaka City University, Osaka, 558-8585, Japan
\linebreak
$^3$ Physics Division, Argonne National Laboratory, Argonne,
Illinois 60439, USA}

\date{\today}
 
\maketitle
\bigskip
\bigskip

\begin{abstract}
A dynamical approach is developed to predict the $\pi N$ scattering 
amplitudes starting with the constituent quark models. 
The first step is to apply a variational method to solve the three-quark
bound state problem. The resulting wave functions are used to 
calculate the $N^* \rightarrow \pi N, \eta N, \pi\Delta$ 
vertex functions by assuming that the $\pi$ and $\eta$ 
mesons couple directly to quarks.
These vertex functions and the predicted baryon bare masses then 
define a Hamiltonian for $\pi N$ reactions. We apply a unitary
transformation method to derive from the constructed Hamiltonian
a multi-channel and multi-resonance reaction model for predicting
the $\pi N$ scattering amplitudes up to $W = 2$ GeV.
With the parameters constrained by
the $\Delta(1232$) excitation, we have examined the extent to which
the $\pi N$ scattering in $S_{11}$ channel can be
described by the constituent quark models based on the 
one-gluon-exchange or one-meson-exchange mechanisms.
It is found that the data seem to favor the
spin-spin interaction due to one-meson-exchange and the tensor
interaction due to one-gluon-exchange. A phenomenological
quark-quark potential has been constructed to reproduce the
$S_{11}$ amplitude. 

\end{abstract}

\bigskip
\bigskip
\bigskip

\bigskip
\bigskip
\newpage
         
\section{INTRODUCTION}

The constituent quark models have long been used to investigate
the structure of nucleon resonances. Most of the earlier 
works\cite{green,feder,chen,dalitz} were based on
phenomenological forms of 
residual quark-quark ($qq$) interactions. 
With the development of Quantum Chromodynamics(QCD), a more fundamental
approach was developed\cite{isgur,caps1,capstick} 
by assuming that the residual
$qq$ interactions can be parameterized as 
the Fermi-Breit form of one-gluon-exchange mechanism\cite{deru}. 
In recent years, an alternative approach has been 
developed\cite{gloz1,gloz2,gloz3} based on
the assumption that the residual $qq$-interaction
is due to the exchange of octet Goldstone bosons. 
With appropriate phenomenological tuning, both approaches
can reproduce the general structure of the 
baryon spectra listed by Particle Data Group (PDG)\cite{pdg}.
Some attempts\cite{valc,dzie,shen} 
have also been made to develop hybrid models including
both one-gluon-exchange and one-meson-exchange quark-quark interactions.
To make progress, it is important to develop an approach to
distinguish all of these constituent quark models 
using 
additional
experimental data.

The main question we want to address in this work is how the  
constituent quark models can be tested against the $\pi N$ scattering data.
It is  common to compare their predicted masses and decay widths with
the data listed by PDG.  All calculations of decay widths have been 
done\cite{copley,foster,koni,li,caps2,iach} 
perturbatively. For example, the width of the 
decay of $N^*$ into a $\pi N$ state is calculated
by evaluating the matrix element $\langle N^*|O(k)|N\rangle $ with an appropriate
operator $O(k)$ describing how pions are coupled to quarks. 
The interactions between the outgoing mesons and baryons are neglected.
It has been found that such a perturbative calculation can at best describe 
the general 
qualitative 
trend of the data, but not the quantitative
details. For example, the widths of $N(1535) \rightarrow \pi N, \eta N,
\pi \Delta$ predicted
by Ref.\cite{caps2} are ($14.7\pm 0.5, 14.6 \pm 0.4, 1.4\pm 0.3$) MeV$^{1/2}$
, which do not seem in quantitative agreement with
the empirical values ($8.0\pm 2.8, 8.1\pm 0.8, 0.$) MeV$^{1/2}$
determined in Ref.\cite{manley}. 

It is important to note here that the PDG's values
are extracted from the experimental $\pi N$ amplitudes which
contain both resonant and non-resonant components. In most partial waves,
the non-resonant mechanisms are important
; one can see this from the fact
that  most of the resonances identified by PDG are in fact
not visible in $\pi N$ and $\gamma N$ cross section data.
By the unitarity condition, therefore the extracted resonance parameters
"inherently" contain non-resonant contributions.  
Furthermore, the separation of non-resonant components from
the full amplitudes is a model-dependent procedure.
The available amplitude 
analyses\cite{manley,hohler,cutkosky,arndt,batinic,vrana}
have yielded very different resonance
parameters in many cases.  
Clearly, except in 
a region where the non-resonant contributions are
negligibly small, the comparison of the 
PDG values (or values from other amplitude analyses) with the 
decay widths calculated perturbatively from the constituent quark models
could be very misleading. In particular, a perturbative calculation
of decay widths is obviously not valid for 
cases in which
two nearby resonances in the same partial wave 
can couple with each other through their coupling with
the meson-nucleon continuum.
Similar precautions must also be taken 
in comparing the predicted masses with the PDG values. These issues
concerning the comparison of the PDG data with the predictions from 
constituent quark models were discussed in Ref.\cite{lee}. 

To have a more direct 
test of constituent quark models, we will explore in
this work a nonperturbative approach that 
takes account of
the unitarity
condition and can relate the $\pi N$ scattering amplitudes directly to
the predicted internal quark wave functions of baryons. Our approach is guided by
a dynamical model of $\pi N$ and $\gamma N$ reactions developed 
in Ref.\cite{satolee} (SL model). It was shown there 
that the $\pi N$ and $\gamma N$ reactions
up to the $\Delta (1232)$ energy region can be described by the following Hamiltonian
\begin{equation}
  \label{SLHam}
  H = H_0 + \Gamma_{\Delta \leftrightarrow \pi N, \gamma N}
  + v_{\pi N, \pi N} + v_{\pi N, \gamma N} \, , 
\end{equation} 
where $\Gamma_{\Delta \rightarrow \pi N, \gamma N}$ describes the 
$\Delta \leftrightarrow \pi N, \gamma N$ transitions,
and $v_{\alpha,\beta}$ are the non-resonant interactions.
It was found that the unitarity condition and the 
non-resonant interaction $v_{\alpha,\beta}$ can shift the
mass of $\Delta$ by about 60 MeV and account for as much as about 40 $\%$ of
the M1 strength of 
the $\Delta \rightarrow \gamma N$ decay. This provides
an explanation of a long-standing discrepancy between the 
M1 value predicted by the constituent
quark model and 
the PDG value.
It is therefore natural to conjecture that
$H_0$ of Eq.\ (\ref{SLHam}) can be identified with the model Hamiltonian of 
a constituent quark model, 
and $\Gamma_{\Delta \rightarrow \pi N, \gamma N}$ correspond to
the decay amplitudes calculated perturbatively using the 
resulting baryon wave functions.
This assumption is then similar to what was used in a dynamical 
study\cite{arima} of 
the
$\pi N$ scattering amplitude in $S_{11}$ 
channel within a constituent quark model.
However, Ref.\cite{arima} did not consider 
the non-resonant interaction $v_{\alpha,\beta}$ and employed 
very simple internal wave functions 
for the
 $N^*(S_{11})$ states.
In this work, we will extend the Hamiltonian Eq.\ (\ref{SLHam}) to consider the multi-channel
and multi-resonance cases. As a first step, we will focus on the
$S_{11}$ channel.
We will concentrate on analyzing the dynamical content of our approach
in this rather complex channel.

There have 
been 
some attempts to
understand the constituent quark models within QCD.
Manohar and Georgi\cite{monohar} 
argued that in the kinematic region between the chiral symmetry
breaking scale $\Lambda_{\chi {\rm SB}} \sim 1$ GeV and the QCD confinement
scale $\Lambda_{\rm QCD} \sim 0.1 - 0.3$ GeV, the effective theory for hadrons
is defined by the 
Lagrangian
\begin{eqnarray}
  \label{QCDL}
  L(g,G,\phi) =\bar{\psi}(i\partial^\mu\gamma_\mu - m_q)\psi
  - i g\bar{\psi}G^\mu \gamma_\mu \psi 
  - \frac{g_A}{2 f} \bar{\psi} \gamma^\mu \gamma_5 \psi \partial_\mu \phi
  +\cdot\cdot\cdot\cdot
\end{eqnarray}
where $m_q$ is the constituent quark mass,  
$\psi$, $G^\mu$, and $\phi$ are the fields for constituent quarks 
, gluons, and Goldstone bosons, respectively. The most crucial dynamical
assumption of this approach is that the
Goldstone bosons are coupled directly to the constituent quarks by the flavor SU(3)
symmetry characterized by the coupling constant $g_A/(2f)$. 
This is consistent with the notion that the Goldstone bosons
result from the spontaneously breaking of the approximate chiral symmetry 
that characterizes
the QCD Lagrangian.

The Lagrangian Eq.\ (\ref{QCDL}) implies that the constituent quarks could interact
with each other through both the exchanges of gluons and Goldstone bosons.
The usual one-gluon-exchange (OGE) and one-meson-exchange (OME)
$qq$-interaction can be calculated from Eq.\ (\ref{QCDL})
using perturbation theory. 
This conjecture is supported by
a recent Lattice QCD calculation\cite{liu}.
In Ref.\cite{liu}, it was found that the mass splitting
between $N$ and $\Delta$ is largely due to the meson-exchange mechanism.
The need of both flavor-independent (i.e. OGE) and
flavor-dependent (i.e. OME) $qq$-interactions is also suggested by the
baryon mass formula determined 
in an algebraic approach\cite{iach}.
In this work, we will take this point of view and
will consider a general constituent quark model which can have both OGE and
OME mechanisms. This model can be reduced to the previously developed 
OGE or OME models in some limits.

The meson-quark couplings were also included in
other hadron models such as the chiral/cloudy bag models\cite{thomas}.
However the Lagrangian Eq.\ (\ref{QCDL}) with $m_q \sim 200-300$ MeV for up and down
quarks is closest to 
a framework within which
one can hope to understand the dynamical origins of the 
most often used 
constituent quark models based on $qq$-potentials.
Perhaps one can apply the dynamical approach developed in this work to also examine
other hadron models. But this is beyond the scope of this paper.

In section II, we present the model Hamiltonian for baryons and 
our method for solving
the three-quark bound state problem. The calculations of meson-baryon-baryon
vertex functions are given in section III. 
A formalism for $\pi N$ reactions
is developed in section IV. In section V, we analyze the dynamical
content of our approach within a simple model. 
The results and discussions are presented in
section VI. Section VII is devoted to conclusions and discussions on
future developments.

\section{INTERNAL STRUCTURE OF BARYONS}
\subsection{Baryon Hamiltonian}
In this exploratory study, we take the simplest, but often used, 
approach 
and
 assume that
the baryon structure can be described in terms of
three nonrelativistic constituent quarks. The model Hamiltonian is of the following
familiar form
\begin{equation}
  h_B = K + V_{\rm conf} + V_{qq}\ .   \label{str-ham}
\end{equation}
The kinetic energy $K$ is defined by
\begin{equation}
  \label{kin}
  K = \sum_{i=1}^3 \frac{\bbox{p}^2_i}{2m_q} + 3m_q -
  \frac{\bbox{P}^2}{6m_q}\ ,
\end{equation}
where $\bbox{p}_i$ is the momentum of the $i$th quark and $\bbox{P}$ is
the center of mass momentum of the three-quark system.
The constituent quark mass $m_q$ is taken 
to be 340 MeV, which is close 
to the value used
for describing the nucleon magnetic moments within the
simple $(0s)^3$ configuration.
The confinement potential $V_{\rm conf}$ is assumed to be 
of the usual linear form 
\begin{eqnarray}
V_{\rm conf} &=& \sum_{i<j} \alpha_c r_{ij}\ ,
\end{eqnarray}
where $r_{ij}=|\bbox{r}_i - \bbox{r}_j|$ and $\alpha_c$ is 
a constant.

For the residual $qq$-interaction $V_{qq}$ in Eq.\ (\ref{str-ham}), 
 we first consider the cases that either the one-gluon-exchange(OGE)
 model or the one-meson-exchange(OME) model is used.
Both models are derived from taking the static limits of the
one-particle-exchange Feynman amplitudes.
For the OGE model, 
following the previous works, we drop its spin-orbit component and retain only
spin-spin and tensor components. The OME model has the same structure
except that it contains a flavor (isospin) dependent factor
$\bbox{\tau}_i\cdot\bbox{\tau}_j$.
We thus consider the following general form of $V_{qq}$
\begin{eqnarray}
  \label{potential}
  V_{qq} &=& \sum_{i < j} \left[\bbox{\sigma}_i \cdot \bbox{\sigma}_j
  V_{\sigma}(r_{ij})
  + \bbox{\sigma}_i \cdot \bbox{\sigma}_j
  \bbox{\tau}_i \cdot \bbox{\tau}_j V_{\sigma \tau}(r_{ij})
  + S_{ij} V_{T}(r_{ij})
  + S_{ij} \bbox{\tau}_i \cdot \bbox{\tau}_j V_{T \tau}(r_{ij})\right]\ ,
\end{eqnarray}
with
\begin{eqnarray}
  S_{ij} = \bbox{\sigma}_i \cdot \hat{r}_{ij}
  \bbox{\sigma}_j \cdot \hat{r}_{ij}
  - \frac{1}{3} \bbox{\sigma}_i \cdot \bbox{\sigma}_j\ .
\end{eqnarray}
Here,  $\bbox{\sigma}_i$ and $\bbox{\tau}_i$ are respectively the spin and 
isospin operators for the $i$th quark. The radial parts of the potentials
in Eq.\ (\ref{potential}) are defined by the momentum space potentials 
\begin{eqnarray}
  \label{vrank0}
V_{i}(r) & = & \int \frac{dq q^2}{2\pi^2} j_0(qr) \tilde{V}_{i}(q)\ 
\end{eqnarray}
for $i=\sigma, \sigma\tau$, and 
\begin{eqnarray}
  \label{vrank2}
V_{i}(r) & = &  \int \frac{dq q^2}{2\pi^2} j_2(qr) \tilde{V}_{i}(q)\ 
\end{eqnarray}
for $i = T, T\tau$.
The differences between the OGE model and OME model are in the
choices of $\tilde{V}_{i}(q)$ for evaluating
Eqs.\ (\ref{vrank0})-(\ref{vrank2}). 

\subsubsection{OGE Model}

The OGE model is obtained by taking
\begin{eqnarray}
  \label{ogeSS}
  \tilde{V}_{\sigma}(q) &=&  -\frac{4\pi \alpha_s}{4}
  \langle \lambda_i \cdot \lambda_j\rangle 
   \frac{1}{6m_q^2} F_{\rm g}(q)\ , \\
  \label{ogeT}
  \tilde{V}_{T}(q) &=&  -\frac{4\pi \alpha_s}{4}
  \langle\lambda_i \cdot \lambda_j\rangle  \frac{1}{4m_q^2} F_{\rm g}(q)\ ,\\
  \label{ogeSSt}
  \tilde{V}_{\sigma \tau}(q) &=& 0\ ,\\
  \label{ogeTt}
  \tilde{V}_{T \tau}(q) &=& 0\ ,
\end{eqnarray}
where $\lambda_i$ is the color SU(3) generator with
\begin{eqnarray}
  \langle\lambda_1 \cdot \lambda_2 \rangle= -\frac{8}{3}\ ,
\end{eqnarray}
for color singlet baryons considered here, and $\alpha_s$ is the quark-gluon coupling constant.
In Eqs.\ (\ref{ogeSS})-(\ref{ogeT}), we have introduced 
a form factor of the form
\begin{eqnarray}
  \label{ogeForm}
  F_{\rm g}(q) = \frac{\Lambda_{\rm g}^2}{\Lambda_{\rm g}^2 + \bbox{q}^2}\ 
\end{eqnarray}
to regularize the interactions at short distances.
This is consistent with the notion that the constituent quarks are not point
particles within an effective theory. This regularization
of the $qq$-potential is essential in obtaining convergent solutions
for the bound state problem defined by the Hamiltonian $h_B$ (Eq.\ (\ref{str-ham})).
If the potentials are not regularized by form factors,
 the ground state energy is 
not bound
from below.

\subsubsection{OME Model}

Since we only consider baryons with strangeness $S=0$, the OME model is
defined only by the exchange 
of $\pi$ and $\eta$ mesons. Because $\eta$ is isospin singlet, it only
contributes to the 
isospin independent parts of the potential. On the other hand, the exchange
of the isovector pion leads to isospin dependent terms.
The resulting OME model is defined by
\begin{eqnarray}
  \label{omeSS}
  \tilde{V}_{\sigma}(q) &=& - \left(\frac{f_{\eta qq}}{m_\eta}\right)^2
  \frac{1}{3}\frac{\bbox{q}^2}{\bbox{q}^2 + m_\eta^2}F_\eta(q)\ ,\\
  \label{omeT}
  \tilde{V}_{T}(q) &=&   \left(\frac{f_{\eta qq}}{m_\eta}\right)^2
  \frac{\bbox{q}^2}{\bbox{q}^2 + m_\eta^2}F_\eta(q)\ ,\\
  \label{omeSSt}
  \tilde{V}_{\sigma \tau}(q) &=& - \left(\frac{f_{\pi qq}}{m_\pi}\right)^2
  \frac{1}{3}\frac{\bbox{q}^2}{\bbox{q}^2 + m_\pi^2}F_\pi(q)\ ,\\
  \label{omeTt}
  \tilde{V}_{T \tau}(q) &=&  \left(\frac{f_{\pi qq}}{m_\pi}\right)^2
  \frac{\bbox{q}^2}{\bbox{q}^2 + m_\pi^2}F_\pi(q)\ ,
\end{eqnarray}
where $f_{Mqq}$ is the meson-quark coupling constant,
$m_M$ denotes the meson mass.
Here, 
as in the OGE model, we 
introduce a form factor 
\begin{eqnarray}
  \label{omeForm}
  F_{M}(q) = \frac{\Lambda^2_M}
  {\Lambda^2_M + \bbox{q}^2} \ 
\end{eqnarray}
to regularize the potentials at short distances.

If we assume that the $\pi NN$ vertex function can be calculated from
the $\pi qq$ interaction (as defined later in section III), 
the $\pi$-quark coupling constant $f_{\pi qq}$ can be related to the
$\pi NN$ coupling constant. Within the naive $(0s)^3$ configuration 
for the nucleon, one finds that
\begin{eqnarray}
  \label{pNNcpl}
  \frac{f_{\pi qq}}{m_\pi} = \frac{3}{5} \frac{g_{\pi NN}}{2M_N}\ ,
\end{eqnarray}
where $M_N$ is the observed mass of the nucleon and we use the empirical value
$g_{\pi NN}^2 / 4\pi = 14$. 
The coupling constant $f_{\eta qq}$ can be calculated from
$f_{\pi qq}$ by using the flavor SU(3) symmetry,
\begin{eqnarray}
  \label{cpl}
  \frac{f_{\eta qq}}{m_\eta} = \frac{1}{\sqrt{3}}
  \frac{f_{\pi qq}}{m_\pi}\ .
\end{eqnarray}
Since there may be SU(3) symmetry breaking mechanisms,
the coupling constant
$f_{\eta qq}$ could be different from its SU(3) value.
We however do not consider this possibility in defining the OME model and
simply use Eq.\ (\ref{cpl}).

\subsection{Solution of three-quark bound state problem}

With the Hamiltonian $h_B$ (Eq.\ (\ref{str-ham})) defined above,
our first task 
is to solve the following three-body bound state problem
in the $\bbox{P}=0$ rest frame of the system
\begin{eqnarray}
  \label{three body eqn}
  h_B | \bwf{B} \rangle = m_B| \bwf{B} \rangle\ ,
\end{eqnarray}
where $|\bwf{B}\rangle$ is the baryon wave function with
the label  $B$ denoting collectively the spin-parity $J^\pi$ and isospin $T$
;
$m_B$ is the mass eigenvalue.
We use the diagonalization method developed in Ref.\cite{ogaito} to solve 
Eq.\ (\ref{three body eqn}).
The basis states for the diagonalization are formed from harmonic oscillator
wave functions.
 This choice has two advantages: (1) By using appropriate
Jacobi coordinates, the center of mass motion can be separated exactly from
the intrinsic wave functions, (2) The resulting wave functions have the desired
S$_3$ symmetry.

In the diagonalization method, the baryon wave function is expanded as
\begin{eqnarray}
  \label{wf expand}
  |\bwf{J^{\pi}T}\rangle = \sum_{i} a^{J^\pi T}_i |J^\pi T;i \rangle ,
\end{eqnarray}
where the basis wave functions are of the following antisymmetrized form:
\begin{eqnarray}
  \label{general wave function}
  |J^{\pi}T;i \rangle = \sum_{\alpha} c_{\alpha}^{i}
  \left[
    | \phi_{NL}^{\alpha_{\rm space}} \rangle
    \otimes
    | \chi_{S}^{\alpha_{\rm spin}} \rangle
  \right]_{(J)}
  \cdot | \eta_T^{\alpha_{\rm flavor}} \rangle \cdot
  | \varphi^{\rm A} \rangle\ .
\end{eqnarray}
Here $\alpha=\{\alpha_{\rm space},\alpha_{\rm spin},\alpha_{\rm flavor}\}$ 
stands for the S$_3$ symmetry of each part of the wave function.
$ | \phi_{NL}^{\alpha_{\rm space}} \rangle$ and 
$ |\chi_{S}^{\alpha_{\rm spin}} \rangle$
are the spatial and spin wave functions with the orbital angular momentum $L$,
quanta of harmonic oscillator $N$, and spin $S$. They 
couple to give  the total angular
momentum $J$ of the baryon. 
$ | \eta_T^{\alpha_{\rm flavor}} \rangle $ and 
$ | \varphi^{\rm A} \rangle$ are the iso-spin and color wave
functions, respectively.
The color wave function $| \varphi^{\rm A}\rangle$ is totally antisymmetric.
By taking appropriate coefficients $c_\alpha^i$,
the basis state $|J^\pi T;i \rangle$ defined by
Eq.\ (\ref{general wave function}) 
is totally antisymmetric.

The coefficients $a^{J^\pi T}_i$ in Eq.\ (\ref{wf expand}) and the mass eigenvalues 
$m_{J^\pi T}$ are obtained from diagonalizing the matrix 
\begin{eqnarray}
  \label{Hij}
  H_{i,j} = \langle  J^\pi T;i | h_B | J^\pi T;j\rangle .
\end{eqnarray}
In practice diagonalization is performed within a limited number of basis states.
Then the solution of Eq.\ (\ref{three body eqn}) 
is a function of the oscillator range parameter $b$. 
We treat it as a variational parameter and find 
$b$ by  imposing the condition:
\begin{eqnarray}
  \label{variational}
  \frac{\partial m_B}{\partial b} = 0\ .
\end{eqnarray}
The basis state is chosen so that the mass eigenvalue  $m_{J^\pi T}$
does not change by further extension of the basis states.
In practice we include the basis states up to 11$\hbar\omega$.

For later discussions, we write down here the lowest basis wave functions for 
$N$, $\Delta$, and $N^*(S_{11})$ (simply called $N^*$ from now on)
\begin{eqnarray}
  \label{smplN}
  | \bwf{N} \rangle &\equiv& | \case{1}{2}^+ \case{1}{2} ; 1 \rangle
  = \frac{1}{\sqrt{2}}
  |\phi_{00}^{\rm S} \rangle
  \left( | \chi_{1/2}^{\rm MS}\rangle | \eta^{\rm MS}_{1/2}  \rangle
  +
  | \chi^{\rm MA}_{1/2} \rangle   |\eta^{\rm MA}_{1/2} \rangle
  \right) | \varphi^{\rm A} \rangle\ ,\\
  \label{smplD}
  | \bwf{\Delta}  \rangle &\equiv& | \case{3}{2}^+ \case{3}{2} ; 1 \rangle
  = 
  | \phi^{\rm S}_{00} \rangle | \chi_{3/2}^{\rm S} \rangle | \eta^{\rm 
  S}_{3/2} \rangle
  | \varphi^{\rm A} \rangle\ ,\\
  | \bwf{N^*_1} S= 1/2 \rangle &\equiv& | \case{1}{2}^- \case{1}{2} ; 1 \rangle
  = \frac{1}{2}
  \left\{ \left[ 
   | \phi_{11}^{\rm MS} \rangle \otimes |\chi_{1/2}^{\rm MS} \rangle
 - | \phi_{11}^{\rm MA} \rangle \otimes |\chi^{\rm MA}_{1/2}\rangle \right]_{(1/2)}
    | \eta_{1/2}^{\rm MS} \rangle 
  \right. \nonumber \\
  \label{smplN*1}
  &&\mbox{}\ \ \ \ \ \ \ \ \ \ \ \ \ \ \  \left. - \left[
   | \phi_{11}^{\rm MS} \rangle \otimes |\chi_{1/2}^{\rm MA} \rangle
 + | \phi_{11}^{\rm MA} \rangle \otimes |\chi^{\rm MS}_{1/2}\rangle \right]_{(1/2)}
    | \eta_{1/2}^{\rm MA} \rangle 
  \right\} | \varphi^{\rm A} \rangle\ , \\
  \label{smplN*2}
  | \bwf{N^*_2} S=3/2 \rangle &\equiv& | \case{1}{2}^- \case{1}{2} ; 2 \rangle
  = \frac{1}{\sqrt{2}}
  \left\{\left[
      | \phi^{\rm MS}_{11} \rangle \otimes | \chi^{\rm S}_{3/2}
  \rangle
  \right]_{(1/2)}
    | \eta^{\rm MS}_{1/2} \rangle + \left[
      | \phi^{\rm MA}_{11} \rangle \otimes | \chi^{\rm S}_{3/2}
  \rangle
  \right]_{(1/2)}
    | \eta^{\rm MA}_{1/2} \rangle
  \right\} |\varphi^{\rm A} \rangle\ ,
\end{eqnarray}
where $\phi^{\rm S}_{NL}$ denotes  a state which is 
totally symmetric with 
respect to the interchange of 
any pair of quarks, and $\phi^{\rm MS(MA)}_{NL}$ denote 
states with mixed symmetries. Similar upper indices
are also used to specify the symmetry properties of the 
spin wave function $| \chi \rangle$ and
isospin wave function $| \eta \rangle$.
The spatial wave function can be explicitly written as
\begin{eqnarray}
  \langle \bbox{r}, \bbox{\rho} | \phi^{\rm S}_{00} \rangle &=&
  [ R_{00}(r) Y_{0}(\hat{r}) \otimes
  R_{00}(\rho) Y_{0}(\hat{\rho}) ]_{(0)}\ , \\
  \langle \bbox{r}, \bbox{\rho} |\phi^{\rm MS}_{11} \rangle &=&
  -[ R_{00}(r) Y_{0}(\hat{r}) \otimes
  R_{01}(\rho) Y_{1}(\hat{\rho}) ]_{(1)}\ , \\
  \langle \bbox{r}, \bbox{\rho} | \phi^{\rm MA}_{11} \rangle &=&
  [ R_{01}(r) Y_{1}(\hat{r}) \otimes
  R_{00}(\rho) Y_{0}(\hat{\rho}) ]_{(1)}\ ,
\end{eqnarray}
where
\begin{eqnarray}
  \bbox{r} &=& \bbox{r}_1 - \bbox{r}_2\ ,\\
  \bbox{\rho} &=& \bbox{r}_3 - \frac{\bbox{r}_1 + \bbox{r}_2}{2}\ ,
\end{eqnarray}
and $R_{nl}$, $Y_{l}$ are the radial wave function and 
spherical harmonics respectively. The spin and isospin wave functions
in Eqs.\ (\ref{smplN})-(\ref{smplN*2}) can be constructed by using the
well known procedure. 

%
%
\section{meson-baryon-baryon vertex functions}
Within the constituent quark model, the decay of a baryon ($B$) 
into a meson-baryon ($M'B'$) state is determined by the matrix
element
\begin{eqnarray}
  \Gamma_{B'M',B}^\dagger({\bbox{k}}) =
  \langle  \bwf{B'};M' | H_{M}({\bbox{k}})  | \bwf{B} \rangle \ ,
\end{eqnarray}
where $\bwf{B}$ is a bound state wave function generated from the above
structure calculation, and $H_{M}({\bbox{k}})$ is an appropriate
operator describing how a meson $M$ with a momentum ${\bbox{k}}$ is  emitted 
or absorbed by constituent quarks. In most of the previous 
works\cite{copley,foster,koni,li,iach}, one assumes
that $H_{M}(\bbox{k})$ is a one-body operator with the parameters determined 
phenomenologically by fitting some of the partial decay widths listed by PDG. 
Calculations have also been done\cite{caps2} 
by using the $^3P_0$ model for $H_M(\bbox{k})$. 

In this work we assume that $H_M(\bbox{k})$ is a one-body
operator which can be derived directly from the effective Lagrangian Eq.\ (\ref{QCDL}) by taking
the nonrelativistic limit of the Feynman amplitude $\bar{u}_{{\bbox{p}}'}
\gamma_5\gamma^\mu k_\mu u_{{\bbox{p}}}$ for the 
$q\leftrightarrow Mq$ transition.
To be consistent with the nonrelativistic treatment
of constituent quarks, we keep only the terms up to the order of $p/m_q$.
In coordinate space, the resulting $q + \pi \rightarrow q$ 
transition operator is
\begin{eqnarray}
H_{\pi  qq}=\frac{i}{\sqrt{(2\pi)^32\omega_\pi}}
\frac{f_{\pi qq}}{m_\pi}\sum_{i=1}^3
e^{i{\bbox{k}}\cdot{\bbox{r}}_i} \tau^{\alpha} \bbox{\sigma}_i \cdot \left[
  \bbox{k}-\frac{\omega_\pi}{2m_q}({\bbox{p}}_i+ {\bbox{p}}'_i) \right] F(k)\ ,
\label{pi-quark-int}
\end{eqnarray}
where $\alpha$ denotes the $z$-component of pion isospin and
$\bbox{p}_i$ ($\bbox{p}_i'$) is  the derivative operator acting on 
the initial (final) baryon wave function; $\bbox{k}$ and
$\omega_\pi=\sqrt{m_\pi^2+\bbox{k}^2}$ are the momentum and energy of pion,
respectively. 
The  operator $H_{\eta qq}$ for the isoscalar 
$\eta$ meson can be obtained from Eq.\ (\ref{pi-quark-int}) by replacing
the label $\pi$ by $\eta$ and dropping the isospin operator $\tau^{\alpha}$.

We note that the operator structure of Eq.\ (\ref{pi-quark-int}) 
is the same as that used in Refs.\cite{koni,iach}.
In Refs.\cite{koni,iach},
however the coefficients in front of each term are treated
as free parameters.
 Here the relative importance between 
these two terms are fixed by the non-relativistic reduction of the 
effective Lagrangian Eq.\ (\ref{QCDL}).
To take into account SU(3) breaking, we will allow the parameter
$f_{\eta qq}$ to deviate from its SU(3) value given by Eq.\ (\ref{cpl}).
Furthermore, we also introduce an additional form factor 
$F(k)$ to account for the
effect due to the finite size of constituent quarks and mesons.
This is consistent with the procedure used above in defining the
$qq$-potential $V_{qq}$. However, the constituent quark form factor
for the interaction Eq.\ (\ref{pi-quark-int}) could 
be different from that for
the effective $qq$-potential, 
since the mesons associated with 
$H_{\pi qq}$ are time-like whereas those associated with $V_{qq}$ are space-like.
The constituent quark form factors used in the meson-quark interactions will be
introduced in the section discussing our results.

With the operator Eq.\ (\ref{pi-quark-int}), 
we find that the $M + B_i \rightarrow B_f$ vertex
function  can be written as 
\begin{eqnarray}
\langle \bwf{B_f}|H_{\pi qq}|\bwf{B_i}; M \rangle
&=&\sqrt{4\pi} 
\sum_{JM}\langle J_iJM_iM|J_fM_f \rangle  
\langle T_i T T_{iz} T_{z}|T_f T_{fz}\rangle  \nonumber \\
&\times& i^JY_{JM}^*(\hat{q}) \Gamma_{B_fB_i}^{JT}(q)
\delta(\bbox{p}_f-\bbox{p}_i -\bbox{k})\ ,
\label{vtx-pqq}
\end{eqnarray}
where $\bbox{p}_i$, $J_i$ and $T_i$ 
($\bbox{p}_f$, $J_f$ and $T_f$) are the momentum,  spin and isospin of the 
initial (final) baryon, respectively.
Their $z$-components are denoted by 
$M_i$ and $T_{iz}$ ($M_f$ and $T_{fz}$). $T$, $T_Z$ and $\bbox{k}$ 
are the isospin and momentum of meson $M$ ($\pi$ or $\eta$);
$\bbox{q}$ is the relative momentum of the initial meson-baryon system. 
In the center of mass
frame (the rest frame of the final baryon $B_f$), we obviously have
the simplification that $\bbox{p}_i= -\bbox{k}$ and $\bbox{q}=\bbox{k}$.
$\Gamma_{B_fB_i}^{JT}(k)$ contains the $k$-dependence of the vertex 
function 
\begin{eqnarray}
\Gamma_{B_fB_i}^{JT}(k)&=&\frac{3}{\sqrt{(2\pi)^32\omega_\pi}}
\frac{f_{M qq}}{m_\pi}\sqrt{4\pi}
\sum_L\sqrt{\frac{2L+1}{2J+1}}(-1)^{J+L}F(k)
\nonumber\\
&&\times
\left\{-(-i)^{L+1-J}\langle L100|J0\rangle \left(1-\frac{\omega_M}{6m_q}
\right)k 
\langle B_f|||j_L(\textstyle{\frac{2}{3}}k\rho)
  [Y_L\otimes\sigma]_{(J)}\tau^{T}|||B_i\rangle\right.
\nonumber\\
&&
\left.\ \ \ \ \ -\frac{\omega_M}{m_q}
\langle B_f|||j_J(\textstyle{\frac{2}{3}}k\rho)[[Y_J\otimes\nabla]_{(L)}
\otimes\sigma]_{(J)}\tau^T|||B_i\rangle
\right\}\ .
\label{q-function}
\end{eqnarray}

If the simple wave functions Eqs.\ (\ref{smplN})-(\ref{smplN*2})
are used to evaluate Eq.(40), we obtain
the following analytic expressions for the
$\pi N \rightarrow N^*_i$ vertex functions
\begin{eqnarray}
\Gamma^{01}_{N^*_1N}(k)&=& 2 \Gamma_{\pi}(k)\ , \\
\Gamma^{01}_{N^*_2N}(q)&=& - \Gamma_{\pi}(k)\ ,
\end{eqnarray}
with
\begin{eqnarray}
\Gamma_{\pi}(k) & = &
\frac{4}{\sqrt{3}b}\sqrt{\frac{4\pi}{(2\pi)^32\omega_\pi}}
\left(\frac{f_{\pi qq}}{m_\pi}\right)
\left[y+\left(-\frac{1}{2}+\frac{y}{6}\right)
\frac{\omega_\pi}{m_q}\right]\exp(-y)F(k),
\label{ns-vertex}
\end{eqnarray}
where $y=b^2k^2/12$. Likewise
the $\eta N \rightarrow N^*_i$ vertex functions
are found to be
\begin{eqnarray}
\Gamma^{00}_{N^*_1N}(k) = & \Gamma^{00}_{N^*_2N}(k) = -\Gamma_{\eta}(k).
\end{eqnarray}
$\Gamma_{\eta}$ can be obtained from $\Gamma_{\pi}$ defined in
Eq.(43) by replacing the
label $\pi$ by $\eta$.
Note that the relative importance of the decay
vertex functions of $N^*_1$ and $N^*_2$ is completely determined by
the differences in their wave functions given in Eqs.\ (30)-(31).

\section{Dynamical model for $\pi N$ reactions}

With the vertex functions
defined by Eq.\ (\ref{vtx-pqq}), we follow the procedures of the SL model to 
develop a dynamical model for $\pi N$ reactions.
The starting Hamiltonian is assumed to be
\begin{eqnarray}
  \label{dynamical-H}
  H = H_0 + H_I\ ,
\end{eqnarray}
where the free Hamiltonian takes the following second-quantization form
\begin{eqnarray}
  \label{h0}
  H_0 = \sum_{B} \int d \bbox{p}\, \varepsilon_{B}(\bbox{p})
  b^\dagger_B(\bbox{p})b_B(\bbox{p})
  +
  \sum_{M} \int d \bbox{k}\, \omega_{M}(\bbox{k})
  a^\dagger_M(\bbox{k}) a_M(\bbox{k})\ .
\end{eqnarray}
Here, $b^\dagger_B$($b_B$) and $a^\dagger_M$($a_M$) are
the creation (annihilation) operators for the
baryons and mesons respectively, and
\begin{eqnarray}
  \label{baryonH0}
  \varepsilon_B (\bbox{p}) &=& \sqrt{m_B^2 + \bbox{p}^2}\ , \\
  \omega_M (\bbox{k}) &=& \sqrt{m_M^2 + \bbox{k}^2}\ . 
\end{eqnarray}
The baryon mass $m_B$  is generated from 
the structure calculation described in section II, while  
we use the experimental value for the meson mass $m_M$. 

The interaction term in Eq.\ (\ref{dynamical-H}) 
is written in terms of the vertex functions defined in section III
\begin{eqnarray}
  \label{int-H}
  H_I = \sum_{BB'M} \int d \bbox{p}d\bbox{p}' d\bbox{k}
      \left[ \langle  \bwf{B'} |H_{Mqq}|\bwf{B}; M \rangle 
        b^\dagger_{B'}(\bbox{p}') b_{B}(\bbox{p}) a_M(\bbox{k}) + \mbox{h.c.}\right].
\end{eqnarray}
The above interaction Hamiltonian is similar to that of 
the SL model, except that the anti-baryon states are absent here.
As discussed in Ref.\cite{satolee}, 
it is a non-trivial many-body problem 
to calculate $\pi N$ reactions with the use of $H_I$.
To obtain a manageable reaction 
theory, we follow Refs.\cite{satolee,Kobasaoh} and apply the unitary
transformation up to the
second order in $H_I$ to derive an effective Hamiltonian. 
The essence of the unitary transformation method 
applied in Ref.\cite{satolee}
is to absorb the unphysical transition $B \rightarrow M' B'$ with 
$m_{B} < m_{B'}+ m_{M'}$ into non-resonant potentials.
The resulting effective Hamiltonian then takes the following form
\begin{eqnarray}
  \label{effhi}
  H_{\rm eff} &=& H_0+ \Gamma + \Gamma^{\dagger} + \hat{v}\ , 
\end{eqnarray}
where $H_0$ is defined in Eq.\ (\ref{h0}). The vertex $\Gamma^\dagger$ 
contains only the physical decay
process $B \rightarrow M'B'$ with $m_{B} > m_{M'} + m_{B'}$
\begin{eqnarray}
  \Gamma^\dagger = \sum_{MBB^\prime}
  \int d \bbox{k} d \bbox{p} d \bbox{p}^\prime\ 
  \langle \bwf{B'};M' | H_{Mqq} |\bwf{B} \rangle
  b^\dagger_{B'}(\bbox{p}') a^\dagger_{M'}(\bbox{k}') b_{B}(\bbox{p})
   \theta(m_{B} - (m_{B'} + m_{M'}))\ . 
\end{eqnarray}
where $\theta(x) = 1(0)$ for $x > 0 ( x< 0)$.
The non-resonant $MB \rightarrow M'B'$ two-body interactions are defined by
\begin{eqnarray} 
  \hat{v} = \sum_{MM^\prime BB^\prime} \int d \bbox{k} d \bbox{k}^\prime
 d \bbox{p} d \bbox{p}^\prime  
   \langle \bwf{B^\prime}; M^\prime  | \hat{v} |  \bwf{B};M \rangle
  a^{\dagger}_{M'}(\bbox{k}^\prime) a_M(\bbox{k})
  b^{\dagger}_{B'}(\bbox{p}^\prime) b_B(\bbox{p}) .
\end{eqnarray} 
By translation invariance, the potential matrix element has the 
following form
\begin{eqnarray}
\langle \bwf{B^\prime}; M'  | \hat{v} | \bwf{B}; M \rangle
=\delta(\bbox{p}^\prime+\bbox{k}^\prime-\bbox{p}-\bbox{k}) 
\langle \bwf{B^\prime}; M^\prime | v | \bwf{B}; M \rangle .
\end{eqnarray}

\begin{figure}
  \centerline{\epsfig{file=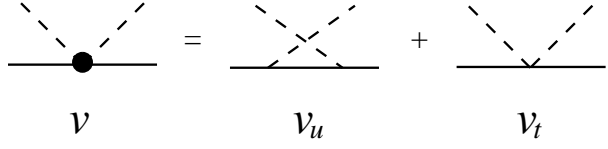,width=8cm}}
  \caption[]{Graphical representation of the non-resonant 
   meson-baryon interaction.}
  \label{fig1}
\end{figure}

To construct the non-resonant interaction $v$, we 
are again guided by the SL model. We first notice that the low energy
$S_{11}$ scattering amplitude can be reproduced very well 
by including the cross nucleon term
and $\rho$-exchange term.
We further notice that the $\rho$-exchange
term in $s$-wave scattering is equivalent to Weinberg's contact term.
Thus for $S_{11}$ scattering considered in this work, it is 
sufficient to consider the non-resonant 
mechanisms illustrated in Fig.\ \ref{fig1}.
However, we need to extend them to include the transitions to
$\eta N$ and $\pi \Delta$ states. 

We write
\begin{eqnarray}
 v = v_t + v_u.
 \label{bgpot}
\end{eqnarray}
We derive the u-channel interaction $v_u$ (the first term in the right-hand 
side of Fig.\ \ref{fig1}) by using
the unitary transformation method 
presented in Ref.\cite{satolee}.
All transitions between $\pi N$, $\eta N$ and
$\pi \Delta$ states are considered. 
The resulting matrix elements of $v_u$ are of
the following form in the center of mass frame:
\begin{eqnarray}
  \label{u-mat}
\langle \bwf{B'};M'|v_u|\bwf{B};M\rangle  & = &
\sum_{B_n}
           \langle \bwf{B'}|H_{Mqq}|\bwf{B_n};M\rangle 
             D(\bbox{k}',\bbox{k})
           \langle \bwf{B_n} ;M'|H_{Mqq}|\bwf{B}\rangle ,
\end{eqnarray}
where $D(\bbox{k}',\bbox{k})$ is given as
\begin{eqnarray}
  \label{u-propag}
D(\bbox{k}',\bbox{k}) & = &
 \frac{1}{2}\left[
\frac{1}{\epsilon_B(\bbox{k})+\omega_M(\bbox{k})
-(\epsilon_{B_n}(\bbox{k}+\bbox{k}')+\omega_M(\bbox{k})+\omega_{M'}(\bbox{k}'))
} \right. \nonumber \\
 & + &
 \left. \frac{1}{\epsilon_{B'}(\bbox{k}')+\omega_{M'}(\bbox{k}')
-(\epsilon_{B_n}(\bbox{k}+\bbox{k}')+\omega_M(\bbox{k})+\omega_{M'}(\bbox{k}'))
} \right]\ .
\end{eqnarray}
For  $\pi N\rightarrow\pi\Delta$ transition with a nucleon intermediate
state, $D(\bbox{k}',\bbox{k})$ takes a different form 
\begin{eqnarray}
  \label{u-propag-piD}
D(\bbox{k}',\bbox{k}) & = &
\frac{1}{\epsilon_N(\bbox{k})+\omega_\pi(\bbox{k})
  -(\epsilon_{N}(\bbox{k}+\bbox{k}')+\omega_\pi(\bbox{k})+
  \omega_{\pi}(\bbox{k}'))
}.
\end{eqnarray}
Here $B$ and $M$ denote the baryon and meson states, respectively.
An intermediate baryon state is denoted by $B_n$.
The allowed intermediate states for each process are listed in Table \ref{table:bgstate}.
The vertex functions in Eq.\ (\ref{u-mat}) can be evaluated by using Eqs.\ (\ref{vtx-pqq})-(\ref{q-function}).
To obtain the partial-wave matrix element from Eq.\ (\ref{u-mat}), we need to
perform the standard angular momentum and iso-spin projections.

For the $\rho$-exchange term $v_t$, we assume that it can be
replaced by the contact term illustrated in Fig.\ 1 and
can be calculated from a contact $ \pi\pi qq$ interaction 
and the nucleon wave functions generated from the 
structure calculations. 
The assumed meson-quark contact interaction is of the  form
\begin{eqnarray}
H_{\rm contact} = \frac{X_t}{4 f_{\pi}^2}(q^{\dagger}\vec{\tau}q)\cdot \vec{\pi}\times
     \dot{\vec{\pi}}.
\end{eqnarray}
Taking the matrix element of  $H_{\rm contact}$,
we obtain 
\begin{eqnarray}
  \label{cont-mat}
\langle \bwf{N};\pi_{i'}|v_t|\bwf{N};\pi_{i}\rangle  & = &
\frac{X_t}{
4f_{\pi}^2(2\pi)^3
\sqrt{4\omega_{\pi}(\bbox{k})\omega_{\pi}(\bbox{k}')}}
 i \epsilon_{ii'k}\tau_k (\omega_{\pi}(\bbox{k})+\omega_{\pi}(\bbox{k}'))
 F_N(\bbox{k}-\bbox{k}')F_t(k)F_t(k').
\end{eqnarray}
Here $F_N(k)$ is the iso-vector form factor of the nucleon completely
determined by the nucleon wave function generated from
the structure calculation described in section II.
$X_t$ and $F_t$ are a phenomenological strength parameter and a 
constituent quark form factor, respectively. These two quantities
can be determined by fitting the $S_{11}$ scattering data 
in the low energy region where the $N^*$ excitation effects are 
negligible. They will be given in section VI.

By using the standard projection operator method\cite{satolee},
it is straightforward to derive 
from the effective Hamiltonian Eq.\ (\ref{effhi})
a calculational framework for $\pi N$ reactions.
The transition operator can be written as
\begin{eqnarray}
  \label{tmat}
  T_{\alpha,\beta} = t_{\alpha,\beta} + 
\sum_{i,j}\tilde{\Gamma}^\dagger_{\alpha,N^*_i}\left[ D^{-1}(E)\right]_{i,j}
\tilde{\Gamma}_{N^*_j,\beta}\ .
\end{eqnarray}
Here $\alpha, \beta$ denote the meson-baryon states $\pi N, \eta N$
and $ \pi \Delta$.
$N^*_i$ are mass eigenstates of Eq.\ (\ref{three body eqn}).
The first term in Eq.\ (\ref{tmat}) is the non-resonant amplitude 
involving only 
the non-resonant interaction $v$
\begin{eqnarray}
  \label{bg-tmat}
  t_{\alpha,\beta}= v_{\alpha,\beta} + \sum_{\gamma} 
v_{\alpha,\gamma}G^0_{\gamma}(E) t_{\gamma\beta}\ ,
\end{eqnarray}
with
\begin{eqnarray}
  \label{bg-propag}
 \left[ G^0_\gamma (E)\right]^{-1} = E - \varepsilon_{B_\gamma}(\bbox{p}) - \omega_{M_\gamma}(\bbox{k})
  + i \epsilon\ .
\end{eqnarray}
The second term in Eq.\ (\ref{tmat}) is the resonant term determined by the dressed $N^*$
propagator and the dressed vertex functions:
\begin{eqnarray}
  \label{nsprop}
  \left[D(E)\right]_{i,j} &=&( E - m_{N^*_i})\delta_{ij}  - \Sigma_{i,j}(E)\ , \\
  \tilde{\Gamma}_{N^*_i,\alpha} &=& \sum_{\gamma}\Gamma_{N^*_i,\gamma}
\left[\delta_{\gamma\alpha} 
+ G^0_{\gamma}(E)t_{\gamma,\alpha}\right]\ , \\
 \tilde{\Gamma}^\dagger_{\alpha,N^*_i} &=& \sum_{\gamma}\left[\delta_{\gamma\alpha}
+t_{\alpha,\gamma}G^0_{\gamma}(E)\right]\Gamma^\dagger_{\gamma,N^*_i}\ .
 \end{eqnarray}
In Eq.\ (\ref{nsprop}), the $N^*$ self-energy is defined by
\begin{eqnarray}
  \label{nsSelf}
  \Sigma_{i,j}(E) = \sum_{\gamma} \Gamma_{N^*_i,\gamma}
 G_{\gamma}^0(E)\tilde{\Gamma}^\dagger_{\gamma, N^*_j}\ .
\end{eqnarray}

\begin{figure}
  \centerline{\epsfig{file=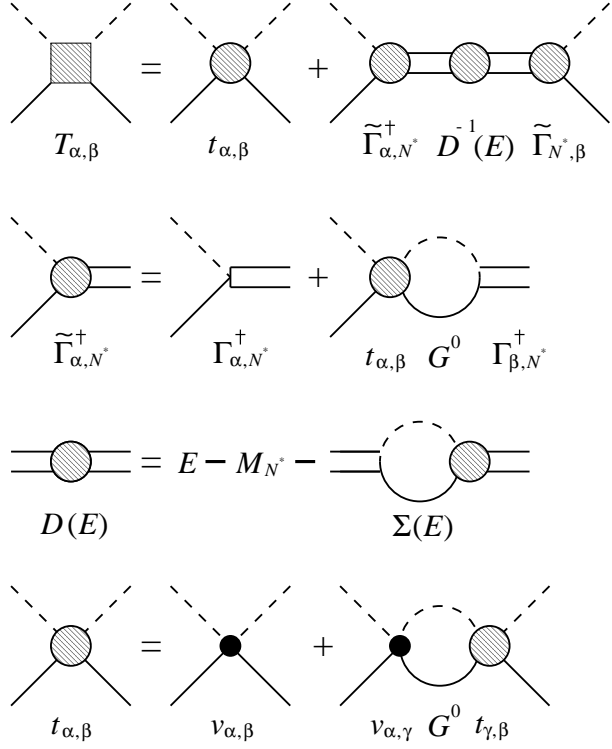,width=8cm}}
  \caption[]{Graphical representation of the scattering equations
    defined by Eqs.\ (\protect\ref{tmat})
    -(\ref{nsSelf}).}
    \label{fig2}
\end{figure}

The scattering equations defined in Eqs.\ (\ref{tmat})-(\ref{nsSelf}) are 
illustrated in Fig.\ \ref{fig2}. They are solved in the 
partial-wave representation using the well-known
numerical method in momentum space. For the $S_{11}$ channel, we consider three
meson-baryon channels $\pi N$, $\eta N$, $\pi \Delta$ and two $N^*$ states.
In the $\pi \Delta$ channel, we account for the width of the
$\Delta$ by modifying the
propagator Eq.\ (\ref{bg-propag}) as
\begin{eqnarray}
 \epsilon_{\Delta}(\bbox{p}) +\epsilon_{M}(\bbox{k})
\rightarrow \epsilon_{\Delta}(\bbox{p})
+ \Sigma_{\Delta}(E-\epsilon_{M}(\bbox{k}))+\epsilon_{M}(\bbox{k})\ ,
\end{eqnarray}
where the $\Delta$ self-energy, $\Sigma_\Delta (\omega)$, is evaluated using
the $\Delta \rightarrow \pi N$ vertex function determined in
Ref.\cite{satolee}.

\section{Results from a simple model}
Within the constituent quark model, the
nature of the effective $qq$-interactions has been investigated
mainly by considering the mass spectrum of nucleon resonances. 
In this work we will apply the reaction model developed in previous
sections to further pin down the effective $qq$-interactions by using 
the $\pi N$ scattering data. To see the merits of this approach, 
it is instructive to
first consider the simplest case in which the $N$, $\Delta$ 
and $N^*$ are described
by the lowest configurations in the harmonic oscillator basis. 
 The spatial 
wave functions for $N$ and $\Delta$
are restricted to $s$-wave. The $N^*$ states in $S_{11}$ channel are
due to $1\hbar\omega$ excitation and hence there are only 
two degenerated states
$| \bwf{N^*_1} S=1/2 \rangle $ and $| \bwf{N^*_2} S=3/2 \rangle $. 
By using these
simple wave functions given 
explicitly in Eqs.\ (\ref{smplN})-(\ref{smplN*2}), we are able to
obtain analytic expressions,
 which facilitates the  understanding of the role of each term in 
the effective $qq$-interactions, Eq.\ (\ref{potential}).
In particular, the flavor (isospin) structure
of the tensor term will be shown to be crucial in determining
the $\pi N$ scattering amplitudes.

With the simple s-wave wave functions Eqs.\ (\ref{smplN})-(\ref{smplD}), 
the mass difference between $N$ and $\Delta$ 
is clearly determined solely by the spin-spin interactions
of Eq.\ (\ref{potential}). It is easy to see
\begin{eqnarray}
  \label{NDsplit}
\delta = m_{\Delta} - m_N = 6 \langle V_{\sigma}\rangle _{s} - 12 \langle V_{\sigma\tau}\rangle _{s}.
\end{eqnarray}
Here $\langle V_{i}\rangle _L$ is the matrix element between
two $qq$-states with relative angular momentum $L=0$ and $1$
\begin{eqnarray}
\langle V_{i}\rangle _L = \int R_{0L}(r) V_i(r) R_{0L}(r) r^2 dr.
\end{eqnarray}
The standard notations $s$ and $p$ are used for $L=0$ and $1$, respectively.
For the OGE model defined by Eqs.\ (\ref{ogeSS})-(\ref{ogeTt}), 
we have $V_{\sigma\tau} =0$ and $\langle V_{\sigma}\rangle_s$ $>$ 0.
For the OME model defined by Eqs.\ (\ref{omeSS})-(\ref{omeTt}), we have $V_{\sigma}=0$ and
$\langle V_{\sigma\tau}\rangle_s$  $<$  0.
From the signs of the coefficients in Eq.\ (\ref{NDsplit}), we can see that
both the OME and OGE models can give a positive $\delta$
and can be tuned to account for the  $\Delta$-$N$ mass
splitting.

For $S_{11}$ states, we diagonalize a $2\times 2$ matrix 
which is obtained by using the
wave functions $| \bwf{N^*_1} S=1/2\rangle $ and 
$| \bwf{N^*_2} S=3/2\rangle $ 
given in Eqs.\ (\ref{smplN*1})-(\ref{smplN*2}) to evaluate Eq.\ (\ref{Hij}). 
Since the spin-spin
interactions in both the OGE and OME models are of  
short-range ($\sim \delta$-function in $r$-space),
we can neglect their matrix elements between $p$-wave relative
wave functions in $\phi^{\rm MS}_{11}$ and $\phi^{\rm MA}_{11}$. 
We then find that the difference between the two
resulting mass eigenvalues has the following analytic form
\begin{eqnarray}
  \label{NSsplit}
\delta^* = m_{N^*_H}- m_{N^*_L} =
\frac{\sqrt{(\delta+\alpha)^2 + 4\alpha^2}}{2},
\end{eqnarray}
where $N^*_L$ and $N^*_H$ denote respectively the lower and higher
mass states. The parameter $\delta$ in Eq.\ (\ref{NSsplit}) has already been fixed by
the $\Delta$-$N$ mass difference in Eq.\ (\ref{NDsplit}). The new parameter 
$\alpha$ is determined by the matrix elements of tensor potentials
between two $p$-wave relative wave functions
\begin{eqnarray}
  \label{alpha}
\alpha = -6\langle V_{T}\rangle _p + 18\langle V_{T\tau}\rangle _p .
\end{eqnarray}
The resulting wave functions for the two $N^*$ states
can be written as 
\begin{eqnarray}
  \label{N*L}
|\bwf{N^*_L}\rangle  & = & \cos\theta |\bwf{N^*_1} S=1/2\rangle  + \sin\theta |\bwf{N^*_2} S=3/2\rangle , \\
  \label{N*H}
|\bwf{N^*_H}\rangle  & = & -\sin\theta |\bwf{N^*_1} S=1/2\rangle  + \cos\theta|\bwf{N^*_2} S=3/2\rangle . 
\end{eqnarray}
The mixing angle $\theta$ also depends
on $\delta$ and $\alpha$ 
\begin{eqnarray}
  \label{mix-angle}
\tan\theta = -\frac{2\alpha}{\alpha + \sqrt{(\delta+\alpha)^2 + 4\alpha^2}}.
\end{eqnarray}

\begin{figure}
  \centerline{\epsfig{file=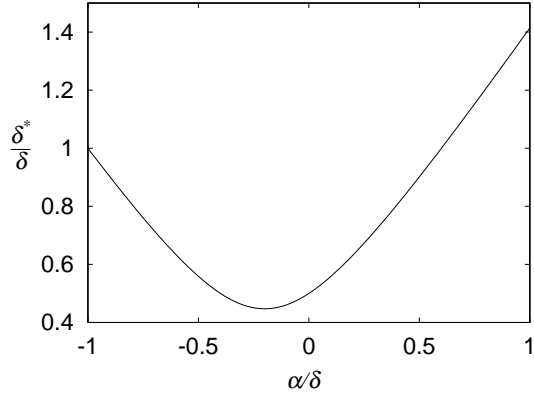,width=7.5cm}}\vspace{5mm}
  \caption[]{Mass splitting $\delta^*$(Eq.(70))
   of $N^*$ as a function of the
  matrix element $\alpha$(Eq.(71) of the tensor potential.
  Both are in units of $\Delta$-$N$ mass difference $\delta$.}
    \label{fig3}
\end{figure}

\begin{figure}
  \centerline{\epsfig{file=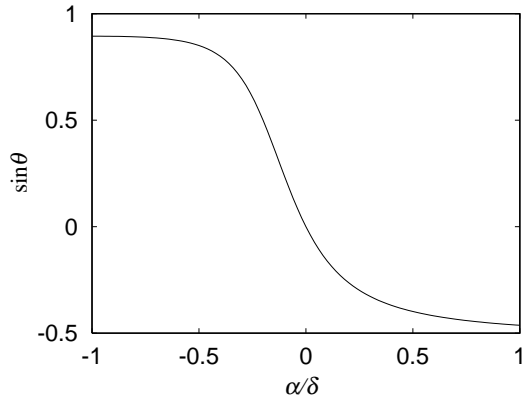,width=7.5cm}}\vspace{5mm}
  \caption[]{Mixing coefficient of the $N^*$ eigenfunctions
  (Eqs.(72)-(73))
   as a function of the matrix element $\alpha$(Eq.(71))
    of the tensor potential.}
    \label{fig4}
\end{figure}

The above expressions indicate explicitly how the structure of $\Delta$
is related to that of  $N^*$  within this simple model. 
For either of the OGE  or OME models, one can
adjust their coupling parameters to fit the same $\Delta$-$N$ 
mass difference $\delta$. But the difference between their 
tensor potentials will lead to very different $\alpha$ (Eq.(\ref{alpha})),
which determines the $N^*$ 
mass splitting $\delta^*$ (Eq.\ (\ref{NSsplit})) and 
wave functions (Eqs.\ (\ref{N*L})-(\ref{N*H})).
We note that the signs of 
$\langle V_T\rangle_p$ evaluated using Eq.\ (\ref{ogeT}) for the OGE model and
$\langle V_{T\tau}\rangle_p$ evaluated using Eq.\ (\ref{omeT}) for the 
OME model are both positive.
It is then clear from Eq.\ (\ref{alpha}) that $\alpha$'s for the OGE model
($V_{T\tau}=0$) and for the OME model ($V_T=0$) are opposite in sign.
Consequently, the phases, defined by Eq.\ (\ref{mix-angle}), of
the $N^*_L$ and $N^*_H$ wave functions will be opposite in sign.
To see this more clearly, we show in Figs.\ \ref{fig3}-\ref{fig4}
the dependences of $\delta^*/\delta$ and mixing coefficient $\sin\theta$
on the parameter $\alpha/\delta$, which measures the strength of the
tensor potential.
In the region $\alpha/\delta < -1$ where the tensor potential
resembles that of the OGE model, the $N^*$ mass splitting can be
smaller than the 
$N$-$\Delta$ mass splitting and the mixing coefficient $\sin\theta$ is
positive.
In the $\alpha/\delta > 0$ region where $qq$-potential 
is close to the OME model, the $N^*$ mass splitting 
$\delta^*$ is most likely
larger than the $N$-$\Delta$ mass splitting $\delta$ 
and the mixing coefficient $\sin\theta$ becomes negative. 
The results shown in Figs.\ \ref{fig3} and \ref{fig4} clearly indicate that the
$N^* $ wave functions depend strongly on the flavor structure of the
tensor potential.
We can make the OGE, OME, or some  mixture of OGE and OME 
models reproduce the same mass splittings $\delta$ and $\delta^*$, but they 
will yield very different $N^*$ 
internal wave functions;
the difference is particularly visible in the
relative phases between the $|\bwf{N^*_1} S=1/2\rangle $ and $|\bwf{N^*_2} S=3/2\rangle $ components.

\begin{figure}
  \centerline{\epsfig{file=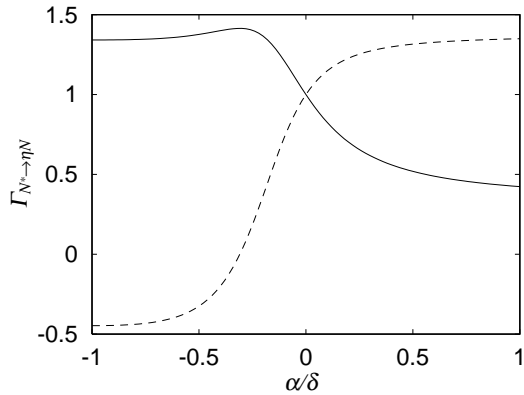,width=7.5cm}}\vspace{5mm}
  \caption[]{$N^* \rightarrow \eta N$ strength as a function of
    the matrix element $\alpha$(Eq.(71)) of the tensor potential.
    The solid and dashed curves are the strengths for
    $N^*_L$ and
    $N^*_H$, respectively.}
    \label{fig5}
\end{figure}

\begin{figure}
  \centerline{\epsfig{file=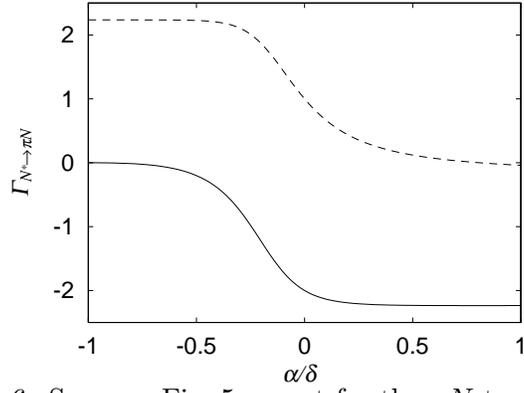,width=7.5cm}}
  \caption[]{Same as Fig.\ \ref{fig5} except for the $\pi N$ transition.}
    \label{fig6}
\end{figure}

We now turn to demonstrating how these different model
wave functions can be distinguished by
investigating the $\pi N$ scattering.
For this discussion, we neglect the non-resonant interaction $v$ of Eq.\ (\ref{effhi})
and the $\pi\Delta$ channel. 
The $\pi N$
scattering amplitude in $S_{11}$ channel is then determined only by the predicted
$N^*$ masses and $N^*\rightarrow \pi N$, $\eta N$ vertex functions. By using 
the vertex functions defined by Eq.\ (\ref{vtx-pqq}) and the 
$N^*$ wave functions Eqs.\ (\ref{smplN*1})-(\ref{smplN*2}), we obtain in the center of
mass frame ($\bbox{p}_f=0, \bbox{p}_i=-\bbox{k}, \bbox{k}=\bbox{q}$)
\begin{eqnarray}
\langle \bwf{N^*_L}|H_{\pi qq}| \bwf{N} ;\pi\rangle  &=& (-2\cos\theta + \sin\theta)\Gamma_{\pi}(k) \\
\langle \bwf{N^*_H}|H_{\pi qq}| \bwf{N} ;\pi\rangle  &=& ( 2\sin\theta + \cos\theta)\Gamma_{\pi}(k) 
\end{eqnarray}
for the $\pi N \rightarrow N^*$ transition, and
\begin{eqnarray}
\langle \bwf{N^*_L}|H_{\eta qq}| \bwf{N} ;\eta\rangle  &=& (\cos\theta  + \sin\theta)\Gamma_{\eta}(k) \\
\langle \bwf{N^*_H}|H_{\eta qq}| \bwf{N} ;\eta\rangle  &=& (-\sin\theta + \cos\theta)\Gamma_{\eta}(k) 
\end{eqnarray}
for the $\eta N \rightarrow N^*$ transition.
Here, $\Gamma_{\pi,\eta}(k)$ are defined by Eq. (\ref{ns-vertex}).
The above equations clearly show that the couplings of $N^*$ states to
$\pi N$ and $\eta N$ continuum are completely dictated by 
 $\theta$, which is related to the strength of tensor potential $\alpha$ via
Eq.\ (\ref{mix-angle}).  This is illustrated in Figs.\ \ref{fig5} and \ref{fig6}. We see that in
the $\alpha/\delta  < 0 $ region where the tensor potential is close
to the flavor independent OGE-type potential ($V_T$), the lower mass 
$N^*_L$ (solid curves) decays mainly 
into the $\eta N$ channel while the higher mass $N^*_H$ (dotted curves) 
favors the decay into the
$\pi N$ channel. The situation is reversed in the $\alpha > 0$ region where
the $qq$-interaction is close to the flavor dependent 
OME model ($V_{T\tau}$).

The results shown in Figs.\ \ref{fig5} and \ref{fig6} suggest that the OGE and OME models
will give very different $\pi N$ scattering amplitudes, even their
parameters can be adjusted to give the same mass splittings
$\delta= m_\Delta-m_N$ and $\delta^*=m_{N^*_H}-m_{N^*_L}$. This
will be seen in our full calculations presented in the next section.

\section{results and discussions}

In this section, we will apply the formulation developed in
section IV to explore to what extent the commonly used
OGE and OME constituent quark models can be consistent with 
the $\pi N$ scattering amplitudes up to 2 GeV.
This is obviously a very difficult task since we need to consider about 20
partial waves which are known to contain resonance excitations.
As a start, we will focus on the $S_{11}$ partial wave.
This partial wave involves strong
coupling between $\pi N$ and $\eta N$ channels and contains
two four-star resonances 
$N(1535)$ and $N(1650)$.
A further complication of this partial wave is that
the position of the first resonance is very close to the
$\eta N$ production threshold ($W_{\rm th} = 1485.7$ MeV). 
A detailed study of this channel is therefore a very useful first step to 
get some insights into our approach.

While we will only focus on the $S_{11}$ channel, the model must be also
consistent with the data associated with the well-studied $\Delta(1236)$
resonance. Here we will use the information from the SL model\cite{satolee} which
is consistent with the present formulation and which 
can describe the $\pi N$ data up to the $\Delta$ excitation.
In our interpretation, the bare mass $m_\Delta = 1300$ MeV and
bare $\Delta \rightarrow \pi N$ form factor determined within the SL model
must be reproduced by our structure calculations.
This is a rather strong constraint on the parameters of the spin-spin parts
of the residual $qq$-interactions and the ranges associated with
the form factors.

Another important ingredient in our investigation is the non-resonant
interaction $v$ of Eq.\ (\ref{bgpot}). 
We demand that the constructed non-resonant interactions
 be consistent with the $S_{11}$ amplitude at low energies where 
the $N^*$ excitation effects are small. This  also provided a 
significant constraint in our
investigation. The $\Delta$ excitation 
and low energy $\pi N$ data were not considered 
in the constituent quark model calculation of $\pi N$ scattering in
Ref.\cite{arima}.

In contrast to usual constituent quark model calculations, the 
determination of the parameters in our approach is a highly nonlinear 
and nonperturbative procedure. For each constituent quark 
model considered, we first carry out extensive structure calculations to
determine the ranges of its parameters
in which the $\Delta$-$N$ mass difference of the SL model can be reproduced.
For each possible set of parameters within thus determined ranges, 
we use the predicted
wave functions for $N$, $\Delta$, and $N^* $  to calculate various
vertex functions that arise from the $\pi qq$ and $\eta qq$ interactions. 
Using these vertex functions, we then calculate
non-resonant potentials.
The scattering equations Eqs.\ (\ref{tmat})-(\ref{nsSelf}) can
then be solved.  The comparison of the predicted $\pi N$ amplitudes
with the data up to about 2 GeV then tells us
whether this set of parameters is acceptable.
This kind of lengthy structure-reaction calculations have to be 
done many times for each considered constituent quark model until the
best fit to the data has been obtained.

In the next few subsections we present our results.

\subsection{Structure calculations}

The baryon mass eigenvalues and wave functions are obtained by diagonalizing
the Hamiltonian $h_B$ defined by Eq.\ (\ref{str-ham}) in the space spanned by the
wave functions defined by Eq.\ (\ref{general wave function}). 
The variational condition Eq.\ (\ref{variational}) 
is imposed to determine
 the oscillator parameter $b$. 
For all of the models considered in this work, we find that it is necessary 
to include configurations up to 11$\hbar \omega$.

\begin{figure}
  \centerline{\epsfig{file=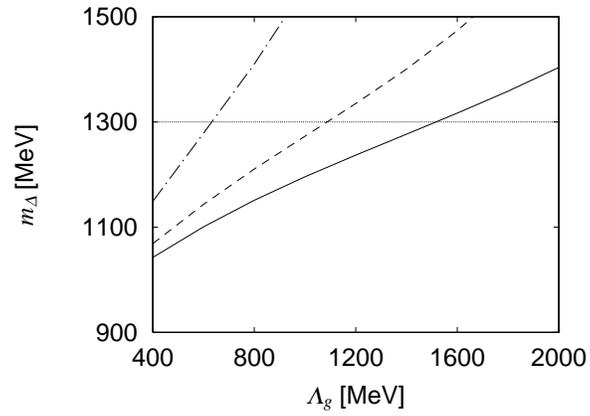,width=8cm}}\vspace{5mm}
  \caption[]{Mass of $\Delta$  as a function of the 
  cutoff $\Lambda_{\rm g}$ in the OGE model. 
  The solid, dashed and dot-dashed curves are
  results with $\alpha_s= 0.8 , 1.0$ and $1.6$, respectively.
  Here the nucleon mass is normalized to $940$ MeV and
  $\alpha_c=4$ fm$^{-2}$ is used.}
    \label{fig7}
\end{figure}

In the one-gluon-exchange (OGE) model defined by 
Eqs.\ (\ref{ogeSS})-(\ref{ogeTt}),  the parameters 
are the vertex cutoff $\Lambda_{\rm g}$, quark-gluon coupling
constant $\alpha_s$, and the strength $\alpha_c$ of a confinement potential.
We find that the $\Delta$-$N$ mass splitting depends strongly
on the parameters $\Lambda_{\rm g}$ and $\alpha_s$. This is illustrated in
Fig.\ \ref{fig7}, where the nucleon mass  is normalized to 940 MeV. 
We see that as $\alpha_s$ is increased from 0.8 (solid curve) to
1.6 (dot-dashed curve),  $m_\Delta=1300$ MeV can be reproduced
only when the cutoff $\Lambda_{\rm g}$ is reduced from about 
1500 MeV to about 600 MeV. Similar results are  obtained also for other 
values of the confinement parameter $\alpha_c$.

\begin{figure}
  \centerline{\epsfig{file=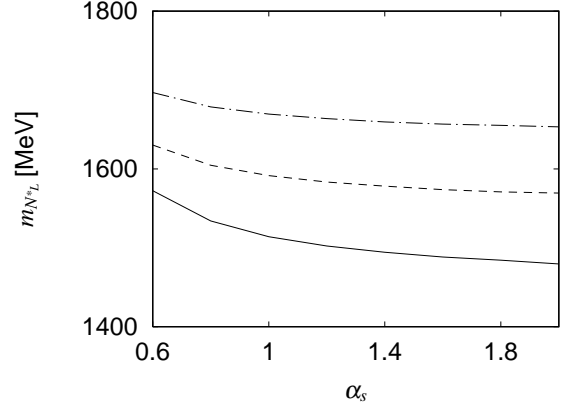,width=8cm}}\vspace{5mm}
  \caption[]{Mass of $N^*_L$ as a function of $\alpha_s$ in the OGE model. 
    The solid, dashed and dot-dashed curves are results with 
    $\alpha_c=3,4$ and $5$ fm$^{-2}$. 
    For each value of $\alpha_c$ and $\alpha_s$, 
    the cutoff parameter $\Lambda_{\rm g}$ is
    determined by requiring the mass of $\Delta$ to be $1300$ MeV.}
    \label{fig8}
\end{figure}

\begin{figure}
  \centerline{\epsfig{file=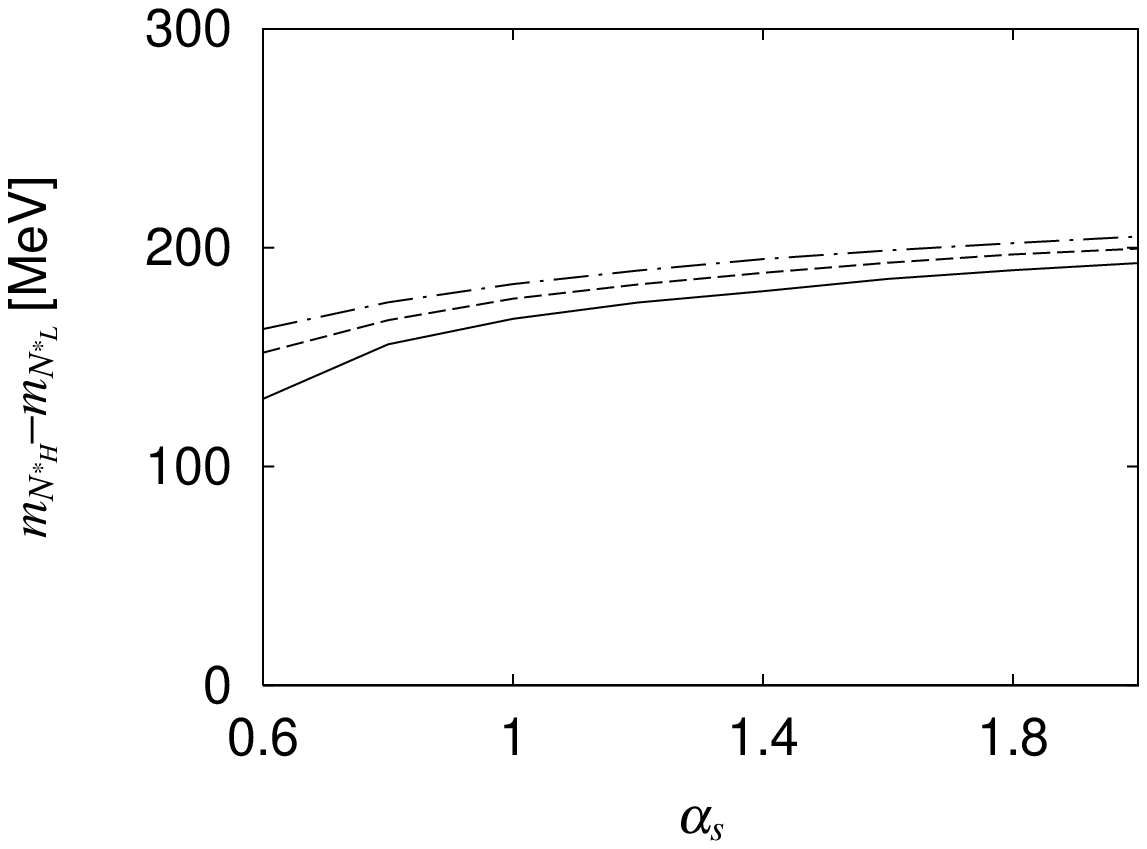,width=8cm}}
  \caption[]{Mass splitting of $N^*$'s in the OGE model. See caption
    of Fig.\ \ref{fig8}.}
    \label{fig9}
\end{figure}

The $N^*$ masses are found to be sensitive to 
$\alpha_c$. In Fig.\ \ref{fig8} we see that for a wide range of 
 $\alpha_s$, the lower mass
$m_{N^*_L}$ can  change by about 200 MeV as $\alpha_c$ is increased from
3 fm$^{-2}$ (solid curve) to 5 fm$^{-2}$ (dot-dashed curve).
However the mass splitting $m_{N^*_H} - m_{N^*_L} \sim 200 $ MeV
is less sensitive to  $\alpha_c$, as 
illustrated in Fig.\ \ref{fig9}.
Since the non-resonant interactions are weak in the $N^*$ region,
we expect that the parameters must be chosen to yield $m_{N^*_L} \sim $ 
1500 - 1600 MeV. From Figs.\ \ref{fig7}-\ref{fig8}, we see that such 
values of  $m_{N^*_L}$ 
and $m_\Delta = 1300$ MeV
can be obtained by choosing $\Lambda_{\rm g} \sim 1100$ MeV, 
$\alpha_c \sim 4$ fm$^{-2}$, and $\alpha_s \sim 1$.

In the one-meson-exchange (OME) model defined by 
Eqs.\ (\ref{omeSS})-(\ref{omeTt}),  the parameters are vertex cutoffs
$\Lambda_\pi$, $\Lambda_\eta$, and the strength $\alpha_c$ of the
confinement potential. 
Here the meson-quark coupling constants $f_{\pi qq}$ and $f_{\eta qq}$ are
fixed by Eqs.\ (\ref{pNNcpl})-(\ref{cpl}) using the standard $\pi NN$ coupling constant
$g_{\pi NN}$ and SU(3) symmetry.  

\begin{figure}
  \centerline{\epsfig{file=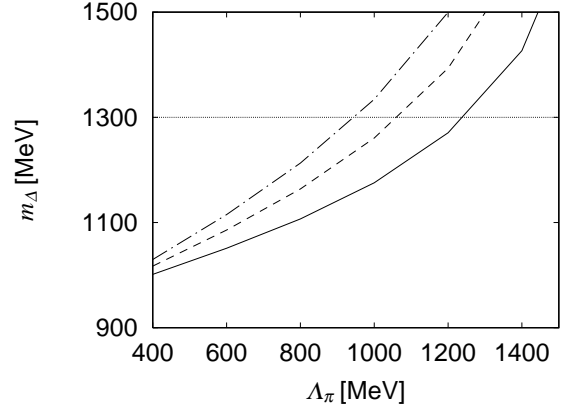,width=8cm}}\vspace{5mm}
  \caption[]{Mass of $\Delta$ as a function of $\Lambda_\pi$
    in the OME model.
    The solid, dashed and dot-dashed curves are the results with
    $\alpha_c= 2,3$ and $4$ fm$^{-2}$ and $\Lambda_\eta = 1$ GeV.}
    \label{fig10}
\end{figure}

\begin{figure}
  \centerline{\epsfig{file=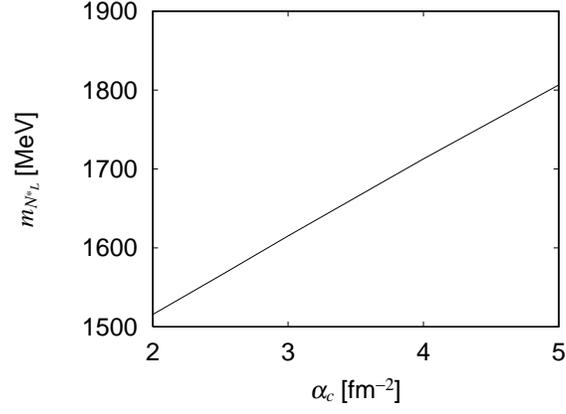,width=8cm}}\vspace{5mm}
  \caption[]{Mass of $N^*_L$ as a function of $\alpha_c$ in the OME model.
    The cutoff $\Lambda_{\pi}$ for each $\alpha_c$ is determined so 
   that the mass of $\Delta$ is $1300$ MeV.}
    \label{fig11}
\end{figure}

\begin{figure}
  \centerline{\epsfig{file=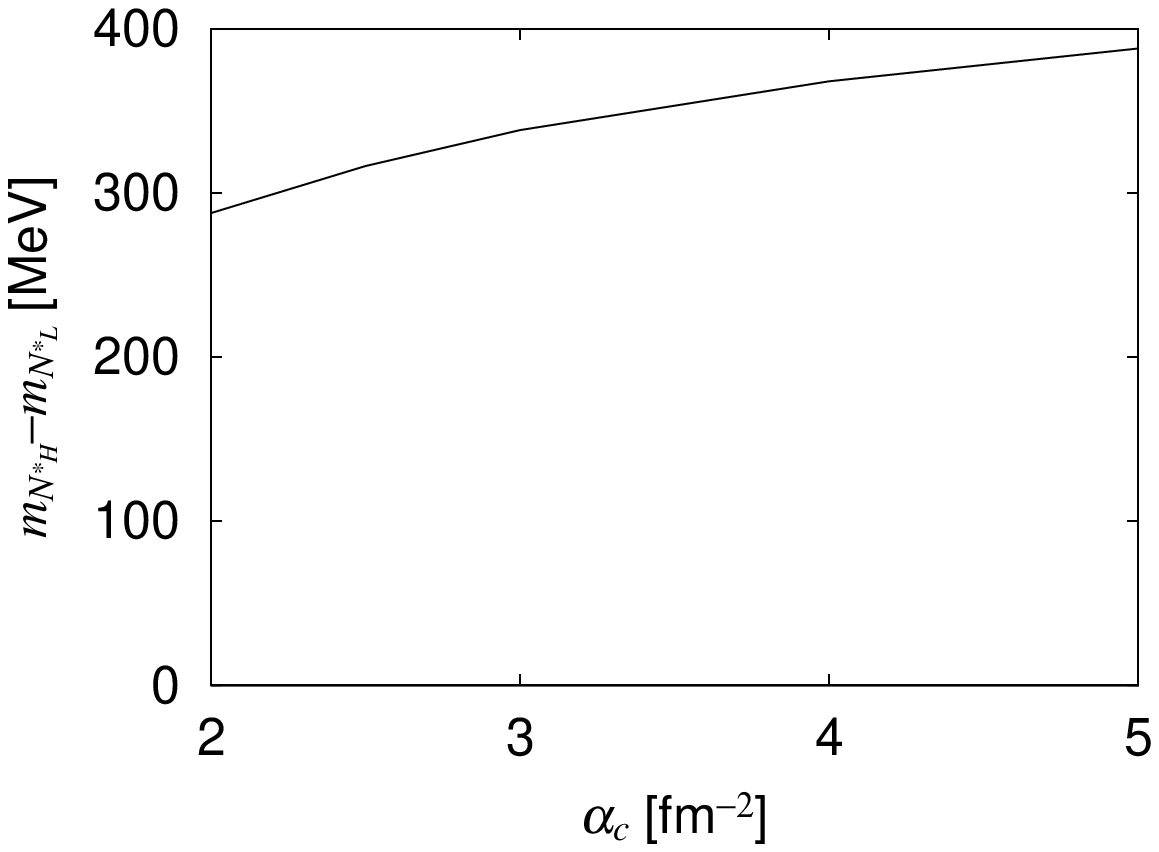,width=8cm}}
  \caption[]{Mass splitting of $N^*$ in the OME model.
    See caption of Fig.\ \ref{fig11}.}
    \label{fig12}
\end{figure}

Since the contribution of $\eta$-exchange potential is rather weak, the
structure calculations are rather insensitive to the range of its cutoff
$\Lambda_\eta$. We set $\Lambda_\eta = 1000$ MeV for simplicity. 
The $\Delta$-$N$
mass splitting is then  determined by the parameters $\Lambda_\pi$ and
$\alpha_c$ alone. 
In Fig.\ \ref{fig10}, we see that $m_\Delta$ is rather sensitive to the
cutoff $\Lambda_\pi$. We also find that the $N^*_L$ mass
is sensitive to  $\alpha_c$, as seen in Fig.\ \ref{fig11}.
The dependence of the mass splitting $m_{N^*_H}-m_{N^*_L}$ on 
$\alpha_c$ is shown in Fig.\ \ref{fig12}. 
We notice here that  $m_{N^*_H} - m_{N^*_L} \sim 300$MeV,
which is about a factor 2 larger than the OGE model value [Fig.\ \ref{fig9}].
This is mainly due to the fact that the OME model has a much stronger
tensor matrix element, as illustrated in Eq.\ (\ref{alpha}) using
the simple model.
To obtain $m_\Delta =1300$ MeV and $m_{N^*_L} \sim 1600 $ MeV, 
we need to use
 $\Lambda_\pi\sim 1100$ MeV and $\alpha_c \sim 2-3$ fm$^{-2}$.

The structure calculations discussed above only identify the possible
ranges of the
parameters for the OGE and OME models. To further pin down the parameters, we now turn to 
discussing our reaction calculations.

\subsection{Reaction calculations}

\begin{figure}
  \centerline{\epsfig{file=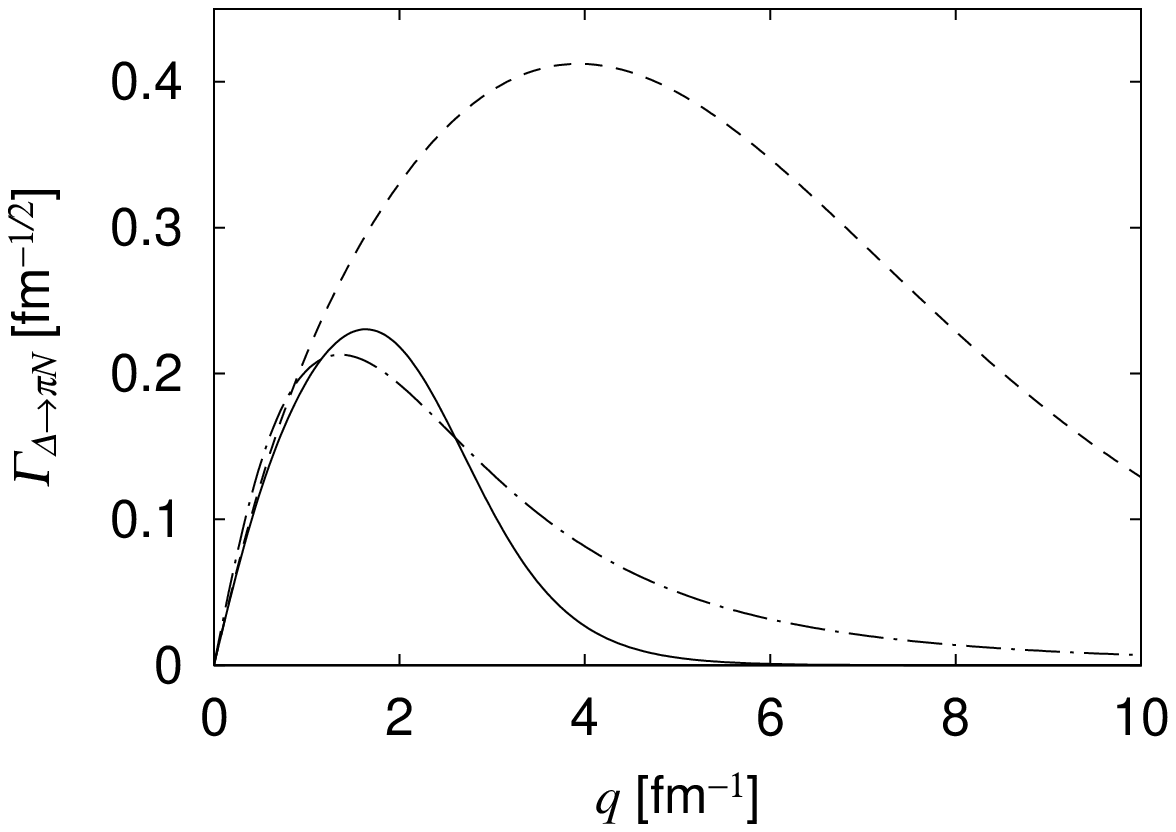,width=8cm}}
  \centerline{(a)}
  \centerline{\epsfig{file=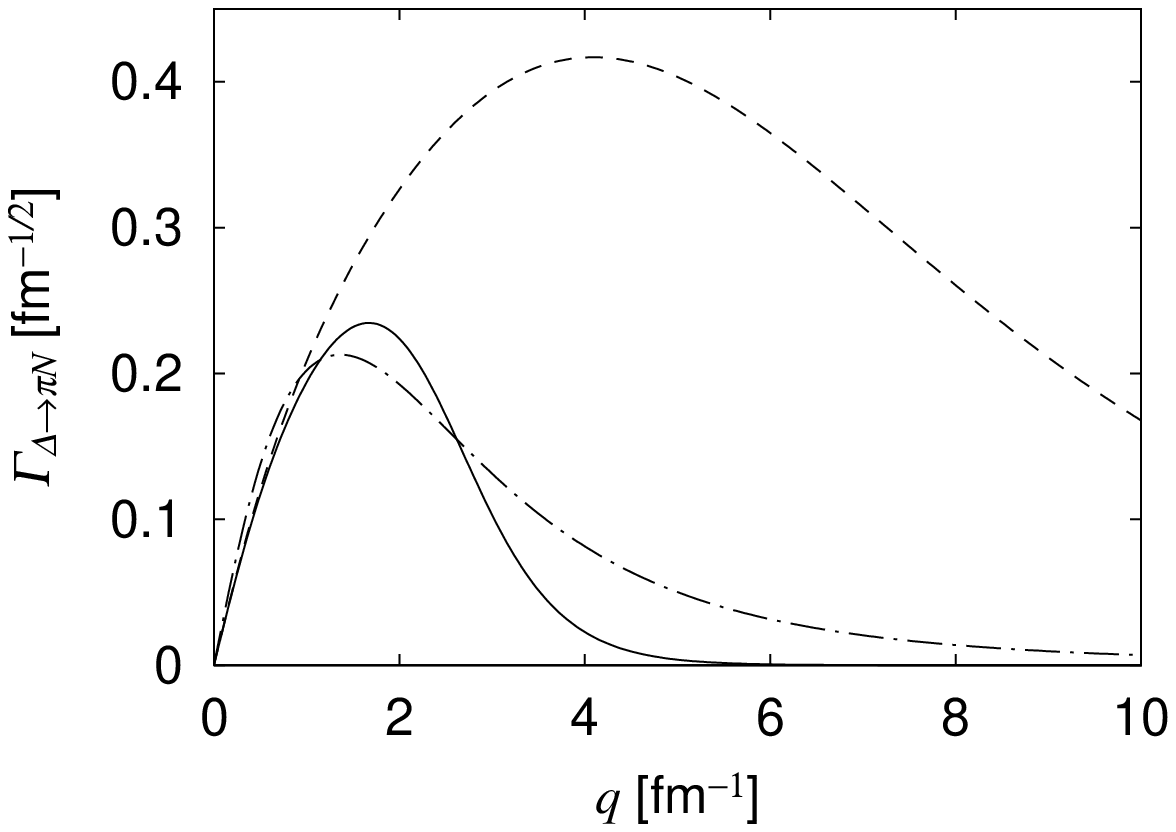,width=8cm}}
  \centerline{(b)}\vspace{5mm}
  \caption[]{$\Delta \rightarrow \pi N$ vertex function in the OGE model 
    (a) and the OME model (b).
    The solid and dashed curves are the vertex functions with and
    without the quark form factor $F(k)$, respectively. The dot-dashed
    curve is the vertex function of the SL model.}
  \label{fig13a}
\end{figure}

As explained in section IV, the first step to perform reaction calculations is 
to use the wave functions
from structure calculations to calculate various vertex functions
using Eqs.\ (\ref{vtx-pqq})-(\ref{q-function}).
We first notice that all of the predicted $\Delta \rightarrow \pi N$
vertex functions
are too hard compared with the bare vertex function of the SL model. 
See the dashed curves in Fig.\ \ref{fig13a}.
Furthermore, no sensible $\pi N$ scattering results can be obtained 
with such hard vertex functions. 
We therefore follow previous works to introduce a constituent quark form
factor to further regularize the predicted vertex functions.
The solid curve in Fig.\ \ref{fig13a}, which is close to the SL model, is
obtained by multiplying the dashed curve by 
the following constituent quark form factor
\begin{eqnarray}
F(k) = \frac{1}{1+e^{(k-k_0)/\Delta_k}}. \label{q-form}
\end{eqnarray}

Another phenomenological aspect in our calculations 
is to allow the $\eta qq$ coupling 
constant to vary around its SU(3) value in calculating
the $\eta$-$BB$ vertex functions.
Namely, we set 
\begin{eqnarray}
\frac{f_{\eta qq}}{m_\eta} = \frac{X_{\eta qq}}{\sqrt{3}}
\frac{f_{\pi qq}}{m_\pi},
\end{eqnarray}
where $X_{\eta qq}$ is a phenomenological parameter 
that is allowed to vary along with
the parameters $k_0$ and $\Delta_k$ [Eq. (\ref{q-form})]
 in fitting the $\pi N$ amplitudes.

The next step  is to fix the non-resonant potential
$v_t$ defined by Eq.\ (\ref{cont-mat}) in the low energy region where the $N^*$
excitation effects are negligible. 
We find that the data up to $W \sim 1300$
MeV can be described well if we 
take $X_t = 0.7$ and set $F_t(k)$ as a dipole form with a 
$500$ MeV cutoff mass. A monopole constituent quark 
form factor with 1 GeV cutoff is also included to soften the
vertex functions of the u-channel interaction $v_u$ (the first term of
Fig.\ \ref{fig1}).

We have found that both the OGE and OME models, 
as defined in this work, can give a good description of the $S_{11}$
amplitudes up to $W <$ about 1500 MeV.
They however can not describe the data at higher energies.
The best results we have obtained are shown in 
Figs.\ \ref{fig14}-\ref{fig15a}.
We see that the OGE model can reproduce the rapid change in phase at
$W \sim 1500$ MeV, while the OME model fails completely. 
The resulting parameters are listed in Tables
 \ref{table:OGEP}-\ref{table:OMEP} for the structure calculations
and in Table \ref{table:Reac} for the calculations of 
$N^* \rightarrow \pi N$, $\eta N$, $\pi \Delta$ form factors.

To understand the differences between the OGE and OME models, 
we show in Figs.\ \ref{fig16a}-\ref{fig18a} the vertex functions calculated
using the best-fit parameters listed in Tables \ref{table:OGEP} and \ref{table:OMEP}.
Generally the spin-flavor structure of $qq$-interaction strongly
influences  the predicted vertex functions.
For the $N^* \rightarrow \pi N$ vertex functions (Fig.\ \ref{fig16a}), $N^*_L$ and
$N^*_H$ have almost the same strength in the OGE model,
while $N^*_L$ decays 
more strongly than $N^*_H$ in the OME model.
This can be understood by simply considering the two main
1$\hbar \omega$ components of $N^*$ given in the first two rows of
Table \ref{table:Coef}.
The main difference between the wave functions of 
the OGE and OME models is in the relative sign between the $S=1/2$ and $S=3/2$
components. This is due to the flavor structure of tensor potential
as illustrated in Fig.\ \ref{fig4} in section V.
This difference in the structure of $N^*$ wave functions also plays an
important role for
the $N^* \rightarrow \eta N$ strength, as seen in Fig.\ \ref{fig17a}.
In the OGE model the strength of $N^*_L \rightarrow \eta N$ is
stronger than $N^*_H \rightarrow \eta N$, while the situation is
opposite in the OME model. For the $N^*\rightarrow \pi\Delta$ vertex
functions compared in Fig.\ \ref{fig18a}, we again see a
 large difference between the two models. The 
OME model predicts a very weak strength for the higher mass $N^*_H$.

In a high energy region $W > 1500$ MeV, we find that the predicted $\pi N$
scattering amplitudes are dominated by the resonant term of
Eq.\ (\ref{tmat}). Thus 
the energy dependence of scattering amplitudes in this region can be 
clearly understood by
examining the $N^*$ propagator $D(E)$ defined by Eq.\ (\ref{nsprop}).
Here we see that the coupling of the
$N^*$ states to $\pi N$, $\eta N$
and $\pi \Delta$ continuum can shift their masses by $\Sigma(E)$.
The poles of $D(E)$ are the resonance positions. These can be
obtained by diagonalizing $D(E)$ in the space
$N^*_L \oplus N^*_H$. The real parts of the resulting
poles include the mass shifts due to the couplings between the two 
$N^*$ states via meson-baryon continuum, and the imaginary parts are the widths of the
resonances. It is important to note here that these effects due to
meson-baryon continuum are not included in the structure calculations
described in section II or any of the existing constituent quark model
calculations. 
In Fig.\ \ref{fig19a} we show the resulting mass eigenvalues of $D(E)$.
The intersections between the mass eigenvalues 
and the dotted lines representing Re$(E^*)=W$
are the resonance positions.
In our fit,  $X_{\eta qq}$ and 
the constituent quark form factors are 
adjusted for each model so  that
the resonance position  for $N^*_L$ lies
around 1535 MeV, the PDG value.
However, the predicted position for  $N^*_H$ 
depend very much on the model.
In the OME model the resonance position of $N^*_H$ is $\sim 1900$
MeV which is clearly too high compared with the PDG value, $1650$ MeV.
Even in the OGE model it is still too high $\sim 1800$ MeV. This is
the main difficulty we have in fitting the data at high energies.
If we choose parameters that fit a lower resonance position for
$N^*_H$, $N^*_L$ becomes too light and
the data below $W < 1500$ MeV can not be fitted at all.

In addition to the resonance positions, 
the OGE and OME models differ also 
in the energy dependence of the imaginary parts of the
self-energy $\Sigma(E)$.
This is illustrated in Fig.\ \ref{fig20a}.
The strong energy dependence of the imaginary part for
$N^*_L$ of the OGE model is due to its stronger coupling with the
$\eta N$ channel, as seen in Fig.\ \ref{fig17a}(a). 
This leads to a very strong 
energy dependence due to the opening of
$\eta N$ threshold; see the solid curve of Fig.\ \ref{fig20a}(a).
The strong coupling of $N^*_L$ to the $\eta N$ channel is 
essential in reproducing the rapid change in 
phase shift around $W=1500$ MeV. This feature can not be generated 
by the OME model within which the $\eta N$ channel is mainly coupled
to the higher mass $N^*_H$, as can be seen
in Fig.\ \ref{fig17a}(b) and \ref{fig20a}(b). 
This is why the OGE model is better than the OME
model in reproducing the data up to about 1550 MeV.

\subsection{Phenomenological model}
The above results  are qualitatively consistent with the
results of the simple model given in section V.
Our findings are perhaps consistent with a recent phenomenological
study\cite{cogeo} of negative parity nucleon resonances. 
It was also found
there that the effective $qq$-interaction is not simply given by
the OME mechanism.
As an attempt to  improve the fit to the $S_{11}$ amplitude, 
we have also explored the mixture of OGE and OME models.
It turns out that such a hybrid model also fails, 
mainly due to the very 
disruptive tensor component of the OME model in determining the
phases of wave functions, as discussed in section V.

We therefore turn to 
investigating a purely phenomenological model. 
We first observe that 
the reason why the OGE model is better than the OME model in
reproducing the main features of the energy-dependence
of the data is that its tensor force yields correct relative
phases between the two $N^*$ wave functions. Consequently, a successful
phenomenological model should have a flavor independent tensor
force similar to that of the OGE model.
 Second, we observe that the difficulty of the OGE model in
reproducing the data at higher energies is mainly due to
its very large mass splitting $m_{N^*_H}-m_{N^*_L}$.
As discussed in section V within the simple model, this problem can
not be fixed within the OGE model unless the 
$N$-$\Delta$ mass splitting is not constrained by the SL model.
This difficulty can be avoided if we have a
short range flavor-dependent spin-spin interaction 
like that of the OME model. 
Qualitatively speaking, the data of $\Delta$ excitation
and $S_{11}$ $\pi N$ scattering seem to 
favor a tensor term due to one-gluon-exchange
and a spin-spin interaction due to one-meson-exchange.

The above considerations have guided us to explore many phenomenological
models. For example, we have found that the $\pi N$ $S_{11}$ amplitudes
can be much better described
by the following phenomenological model
\begin{eqnarray}
\tilde{V}_\sigma &=& 0, \\
\tilde{V}_{\sigma\tau}&=&\frac{4\pi\alpha_{\sigma\tau}}{6m^2_q}F_{\rm ph}(q), \\
\tilde{V}_T&=&\frac{4\pi\alpha_T}{4m_q^2}F_{\rm ph}(q), \\
\tilde{V}_{T\tau}&=&0.
\end{eqnarray}
The resulting parameters are listed in Table \ref{table:PHEN}. 
It is interesting to first
compare the $N^*$ wave functions of this model 
with those of the OGE model.
We see in Table \ref{table:Coef} that 
the  relative phases between the $S=1/2$ and $S=3/2$ components 
in the  $N^*$ wave functions are the same in the two models. 
We therefore expect that
the $N^* \rightarrow \pi N, \eta N, \pi \Delta$ vertex functions also must be
qualitatively very similar. However, we see from Table \ref{table:Coef} that
the phenomenological model apparently
has a stronger tensor potential in mixing the $S=1/2$ and $S=3/2$ components.
As discussed in section V using the simple model, we therefore
expect that the differences between the $N^*_L$ and $N^*_H$
vertex functions must be 
larger than that of the OGE model.
This is exactly what we see by comparing the
results in Fig.\ \ref{fig21a} and Figs.\ \ref{fig16a}(a), \ref{fig17a}(a),
and \ref{fig18a}(a).

According to  Tables \ref{table:OGEP} and \ref{table:PHEN},
 $m_{N^*_H} = 1719.5$ MeV 
for the phenomenological model is lower than the
value 1772.4 MeV of the OGE model. This about 50 MeV shift of $m_{N^*_H}$ is also crucial
in improving the fit at higher energies.

The results from the phenomenological model are compared with the $\pi N$
data in Figs.\ \ref{fig22}-\ref{fig23}.
The predicted $\pi N$ $S_{11}$ phase shifts and $S_{11}$ amplitudes
become closer to the data
than the OGE model.
Moreover  the rapid energy-dependent behavior of the phase shift
around $W \sim 1500$ MeV is reproduced.
This behavior is closely related to the sharp structure in the width of
the $N^*_L$ in the solid curve of Fig.\ \ref{fig24a}(b). 
We also observe that
Fig.\ \ref{fig24a}(b) indicates
that below the $\eta N $ threshold $\sim 1490$ MeV, the imaginary
part of $D(E)$ for 
$N^*_H$ is smaller than that of $N^*_L$. The situation is opposite in the
OME model (Fig.\ \ref{fig20a}(a)). 
This change in the widths is also instrumental in getting better results
at energies near $\eta N$ threshold.

The improvement due to the $N^*$ mass splitting is clearly reflected
in the $\pi N$ scattering amplitudes in Fig.\ \ref{fig23}.
Compared with the OGE model results (solid curves in Fig.\ \ref{fig15a}), the peak positions 
are much better reproduced. 
However, there are still significant discrepancies near the
second peak. This can be understood from the pole positions of T-matrix,
displayed in Fig.\ \ref{fig24a}.
The position of the second resonance
is at about 1700 MeV, still 50MeV higher than 
the PDG value, 1650 MeV. On the other hand,
the first resonance position is very close to the PDG value, 1535 MeV.

Finally the
total cross sections of the $\pi^- + p \rightarrow \eta + n$ reaction
predicted by using our
three models (OME, OGE and phenomenological) are compared with data in Fig. \ref{fig25}.
Near the $\eta N$ production threshold, where $N^*_L$ is expected to 
dominate the cross section, both the OGE and phenomenological 
models explain the data well, while the results from the OME model
are too small due to the weak coupling of its 
$N^*_L$ to the $\eta N$ channel.
The phenomenological model seems to give the best description
of $both$ the $\pi N$ amplitudes and the $\eta$ production cross 
sections. The discrepancy with the data at higher energies in 
Fig.\ \ref{fig25} is mainly due to the
neglect of non-$S_{11}$ partial waves.

Our results in this section suggest that the residual $qq$-interactions
within the constituent quark model are much more complicated than the conventional
OGE and OME models. Within the effective
theory defined by the Lagrangian Eq.\ (\ref{QCDL}), higher order exchanges of 
mesons and gluons
must be considered. The phenomenological model we have obtained is very suggestive.
It remains to be seen whether this model is also consistent with
the data of other partial waves.

\section{Conclusions}

We have developed a dynamical approach to predict $\pi N$ 
scattering amplitudes starting with the constituent quark models.
This is an extension of the dynamical model developed in Ref.\cite{satolee} to
account for the multi-channel and multi-resonance cases.
In this exploratory investigation, we focus on the
$\pi N$ amplitude in the $S_{11}$ channel and only 
consider the most frequently used nonrelativistic
constituent quark models based on either the one-gluon-exchange (OGE)
or the one-meson-exchange (OME) quark-quark residual interactions. 

The first step of our calculations is to choose
appropriate parameters of
the considered constituent quark models to reproduce the bare
parameters associated with the $\Delta$ within 
the $\pi N$ model developed in Ref.\cite{satolee}. 
Here, we apply a variational method to solve the three-quark
bound state problem. The resulting wave functions are then used to 
calculate the $N^* \rightarrow \pi N, \eta N, \pi\Delta$ 
vertex functions by assuming that the $\pi$ and $\eta$ 
mesons couple directly to quarks.
These vertex functions and the predicted baryon bare masses then 
define a Hamiltonian for $\pi N$ reactions. We apply the unitary
transformation method of Ref.\cite{satolee} to solve the $\pi N$ scattering
problem. The final parameters of the considered constituent quark models
are determined by fitting the $\pi N$ scattering data.

\begin{figure}
  \centerline{\epsfig{file=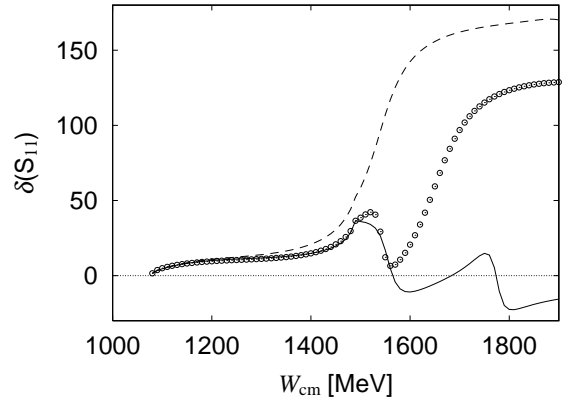,width=8cm}}\vspace{5mm}
  \caption[]{
    Phase shifts of the $\pi N$ scattering in $S_{11}$ channel.
    The solid and dashed curves are the results from
     the OGE and OME models, respectively. 
     Open circles are the data of VPI SP98 \cite{arndt}.}
    \label{fig14}
\end{figure}

\begin{figure}
  \centerline{\epsfig{file=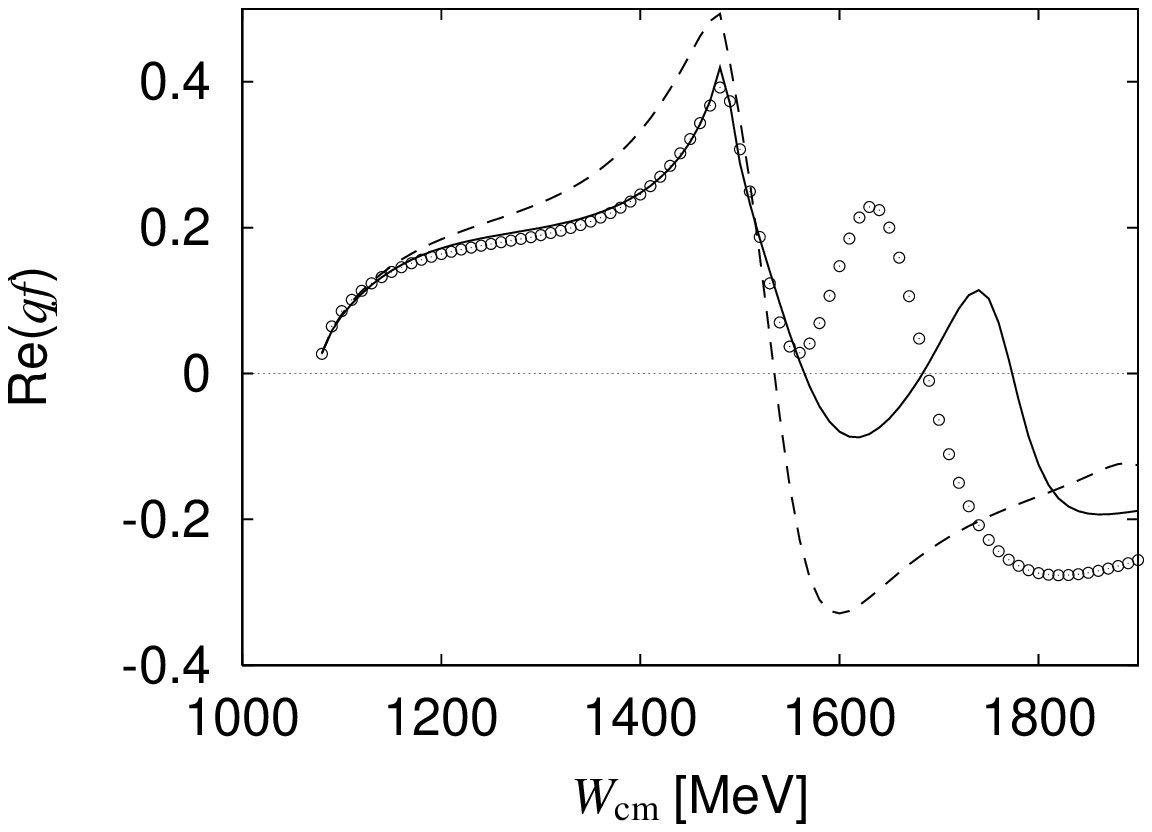,width=8cm}}
  \centerline{(a)}
  \centerline{\epsfig{file=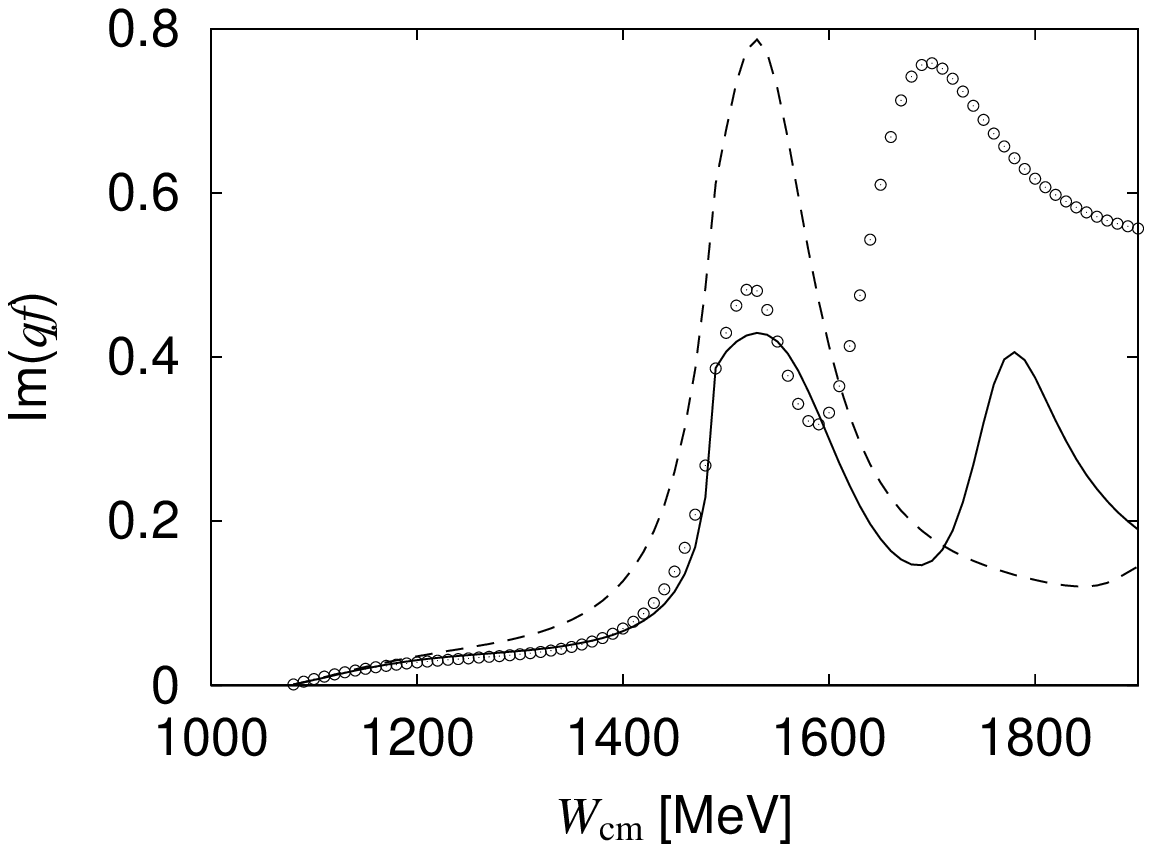,width=8cm}}
  \centerline{(b)}\vspace{5mm}
  \caption[]{Same as Fig.\ \ref{fig14} except for the
real parts (a) and the imaginary parts (b) of the
    $\pi N$ scattering amplitudes in $S_{11}$ channel.}
    \label{fig15a}
\end{figure}

\begin{figure}
  \centerline{\epsfig{file=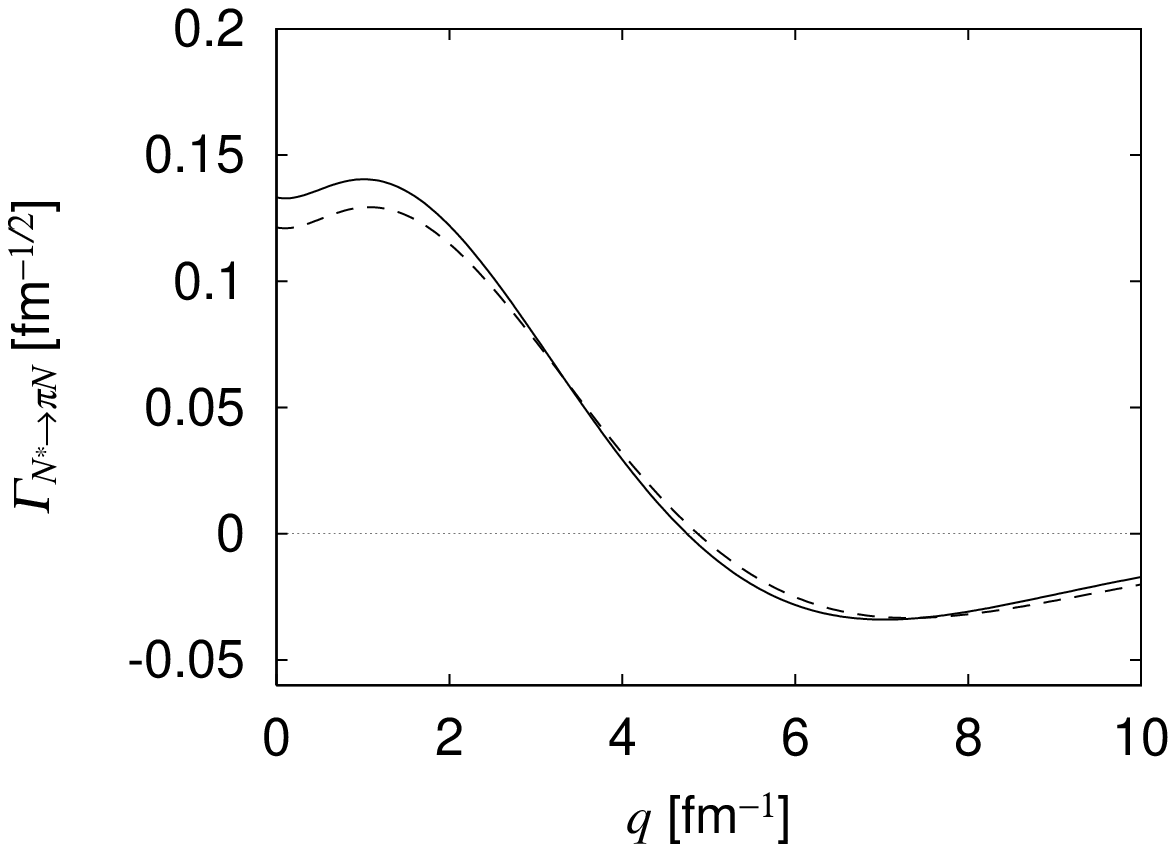,width=8cm}}
  \centerline{(a)}
  \centerline{\epsfig{file=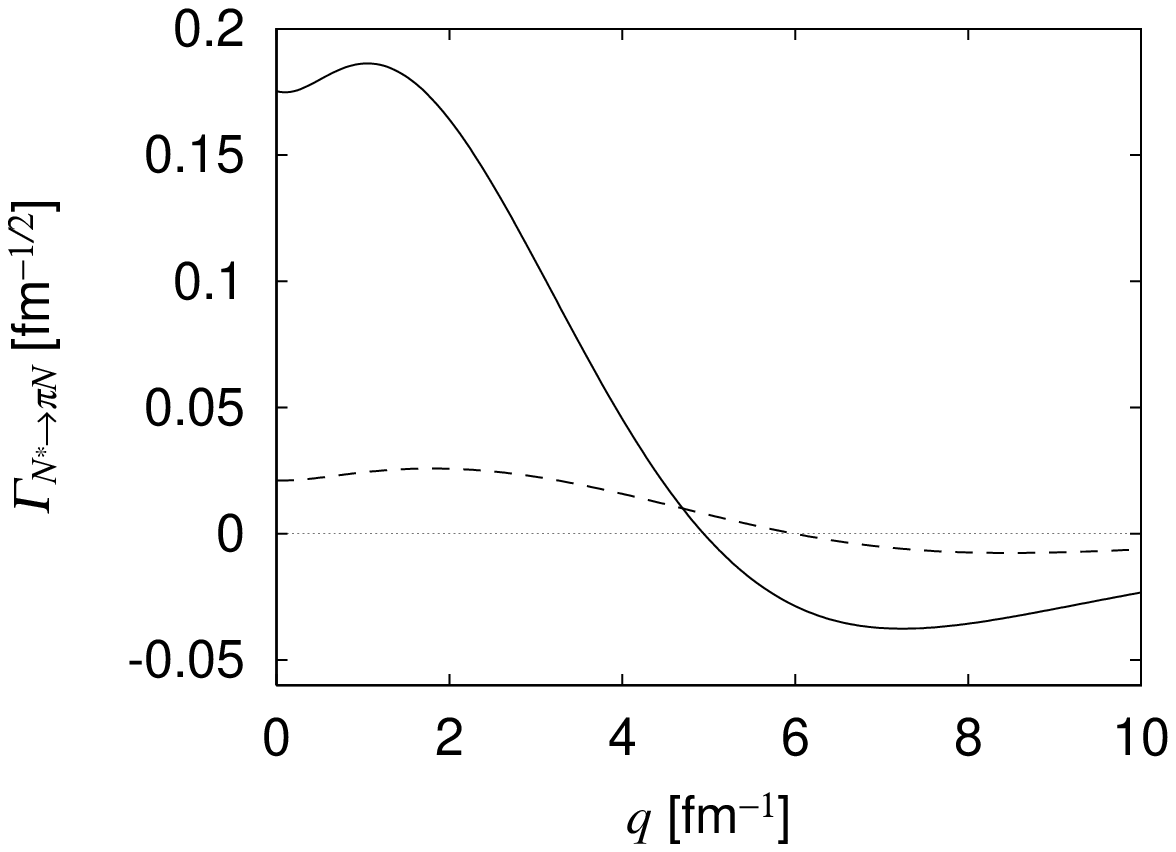,width=8cm}}
  \centerline{(b)}\vspace{5mm}
  \caption[]{$N^* \rightarrow \pi N$ vertex functions calculated
    from the OGE model
    (a) and the OME model (b).
    The solid and dashed curves are for $N^*_L$
    and $N^*_H$, respectively.}
    \label{fig16a}
\end{figure}

\begin{figure}
  \centerline{\epsfig{file=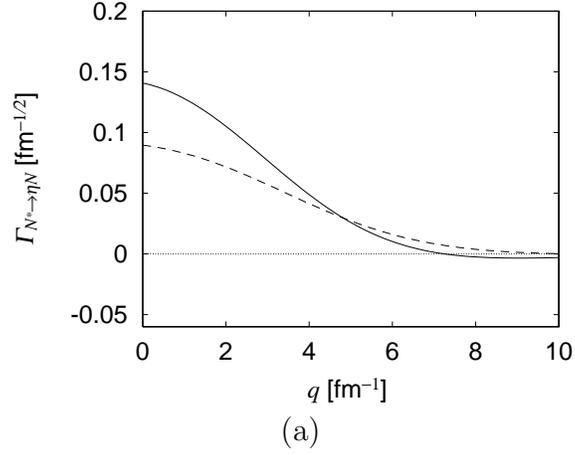,width=8cm}}
  \centerline{(a)}
  \centerline{\epsfig{file=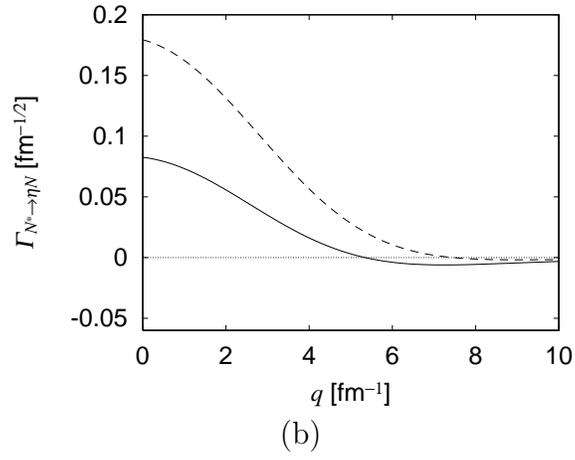,width=8cm}}
  \centerline{(b)}\vspace{5mm}
  \caption[]{Same as Fig.\ \ref{fig16a} except for the
   $N^* \rightarrow \eta N$ vertex functions.}
    \label{fig17a}
\end{figure}

\begin{figure}
  \centerline{\epsfig{file=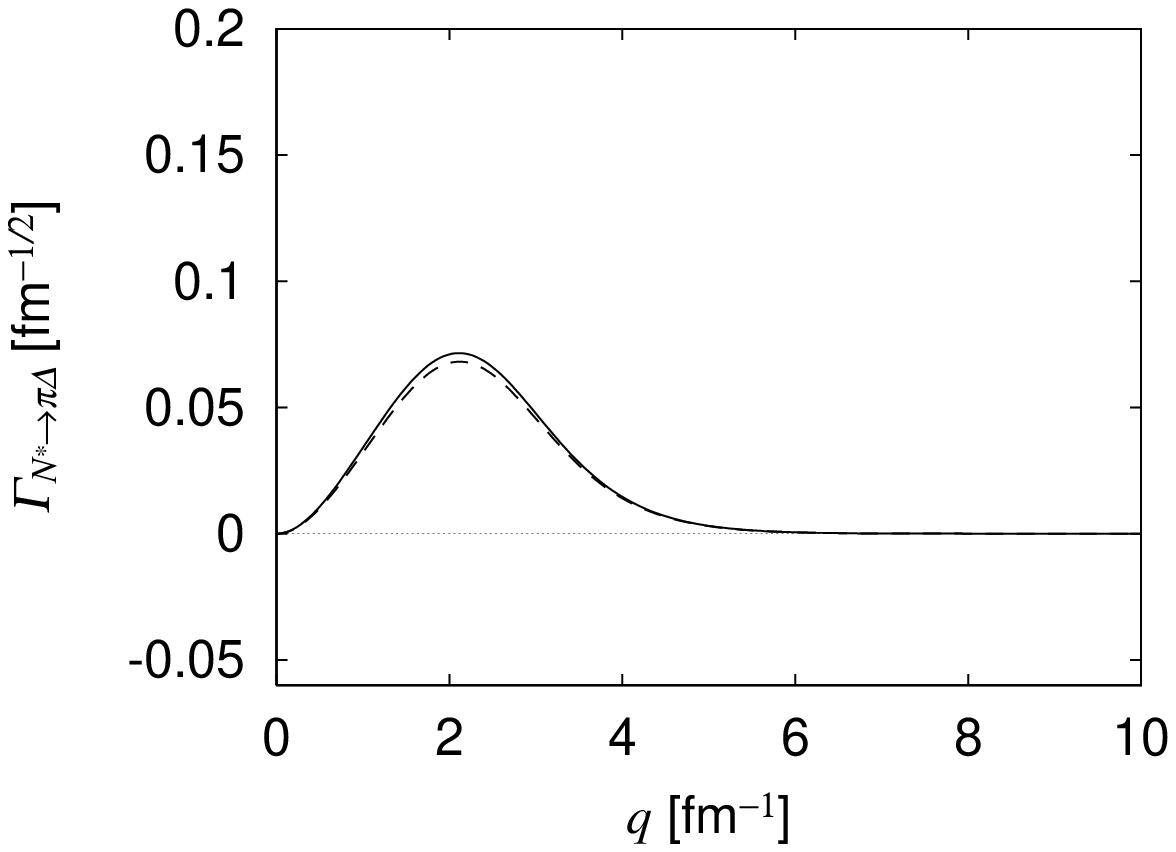,width=8cm}}
  \centerline{(a)}
  \centerline{\epsfig{file=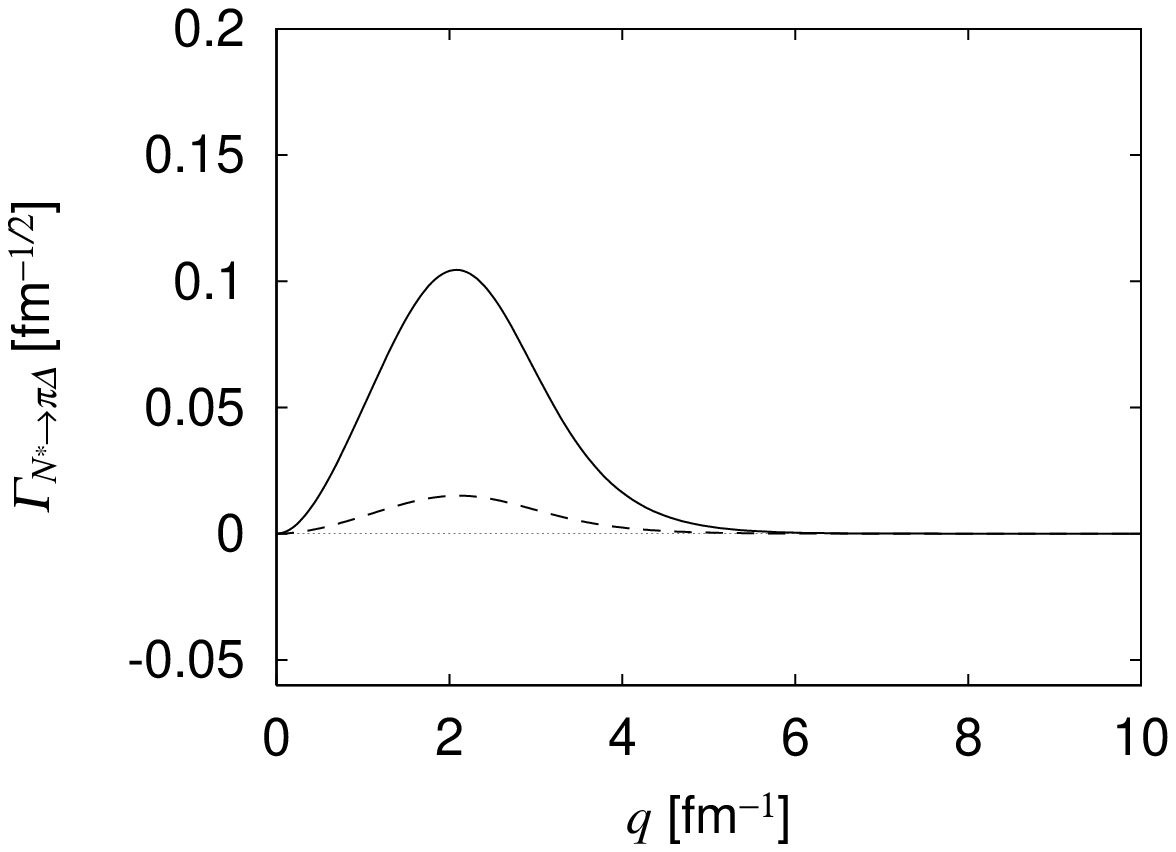,width=8cm}}
  \centerline{(b)}\vspace{5mm}
  \caption[]{Same as Fig.\ \ref{fig16a} except for the
    $N^* \rightarrow \pi \Delta$ vertex functions.}
  \label{fig18a}
\end{figure}

\begin{figure}
  \centerline{\epsfig{file=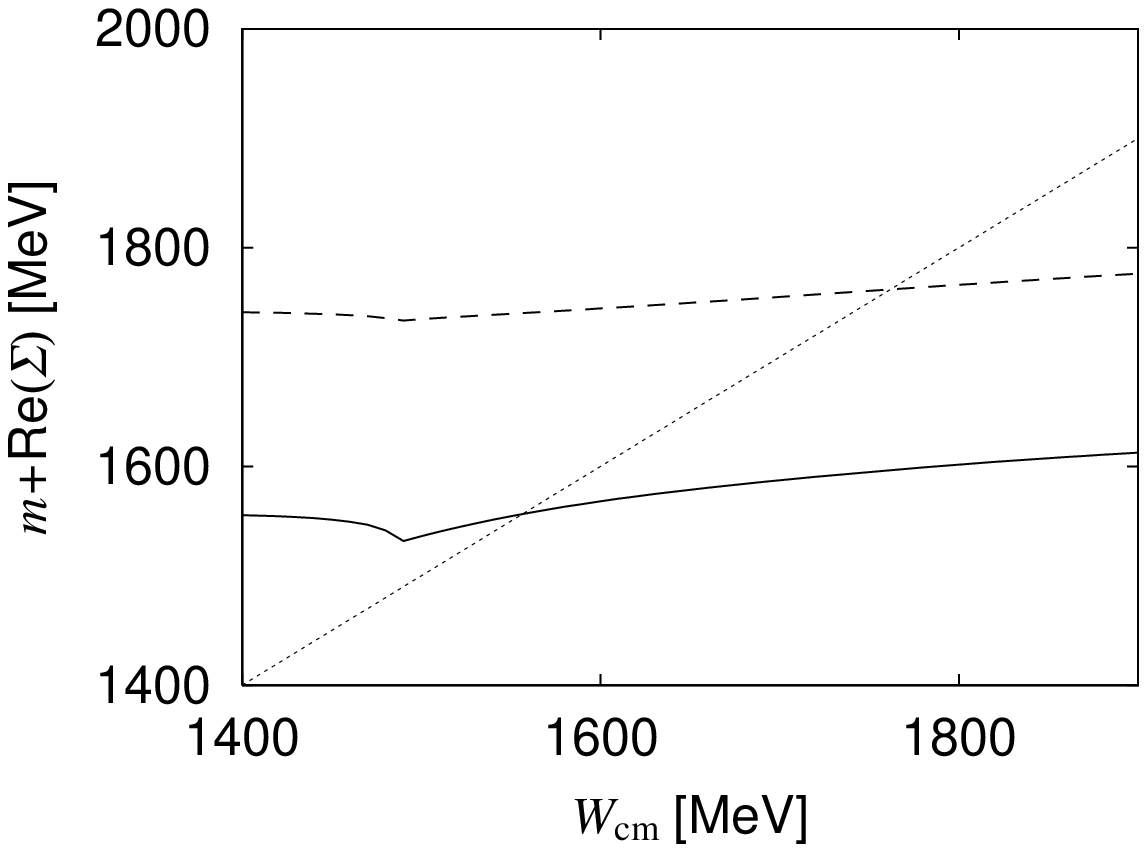,width=8cm}}
  \centerline{(a)}
  \centerline{\epsfig{file=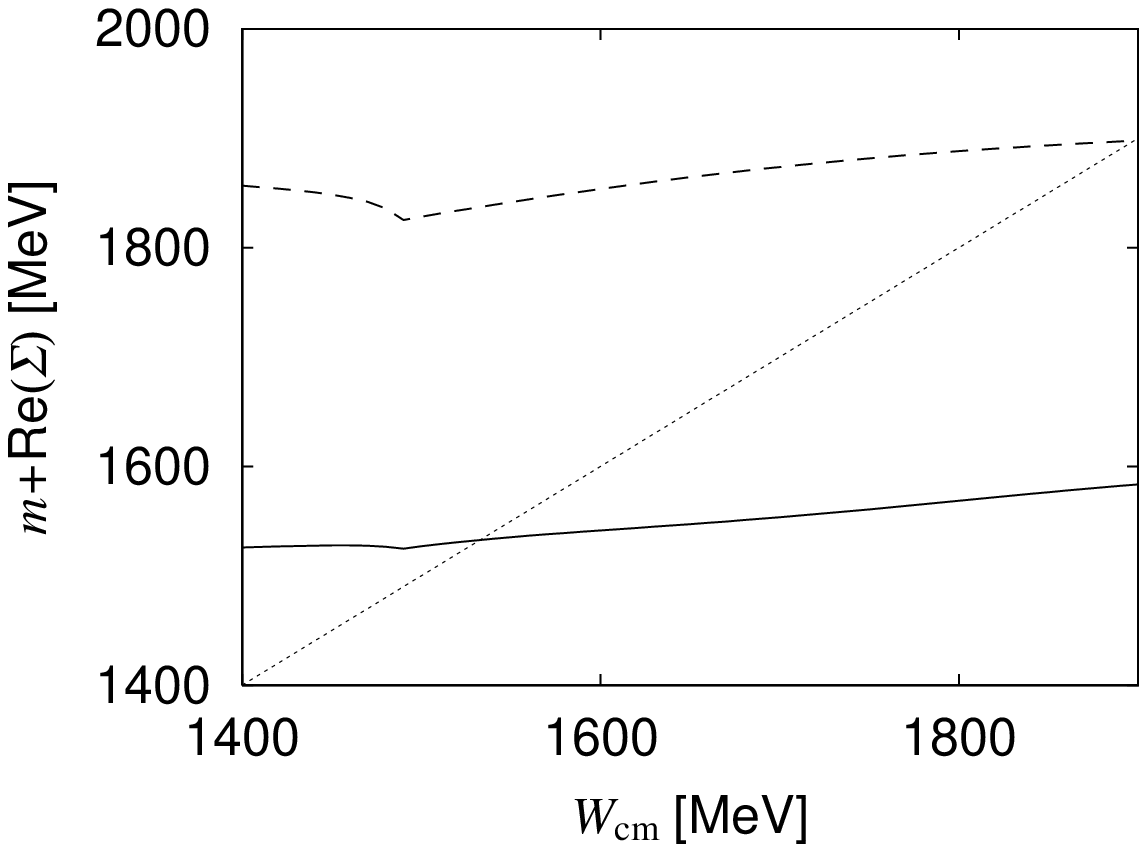,width=8cm}}
  \centerline{(b)}\vspace{5mm}
  \caption[]{Real parts of the eigenvalues of $D(E)$(Eq.(63))
   calculated from the OGE model
    (a) and the OME model (b).
    The solid and dashed curves correspond to the
    masses for $N^*_L$ and $N^*_H$, respectively.
    The dotted line represents Re$(E^*)=W_{\rm cm}$. }
    \label{fig19a}
\end{figure}

\begin{figure}
  \centerline{\epsfig{file=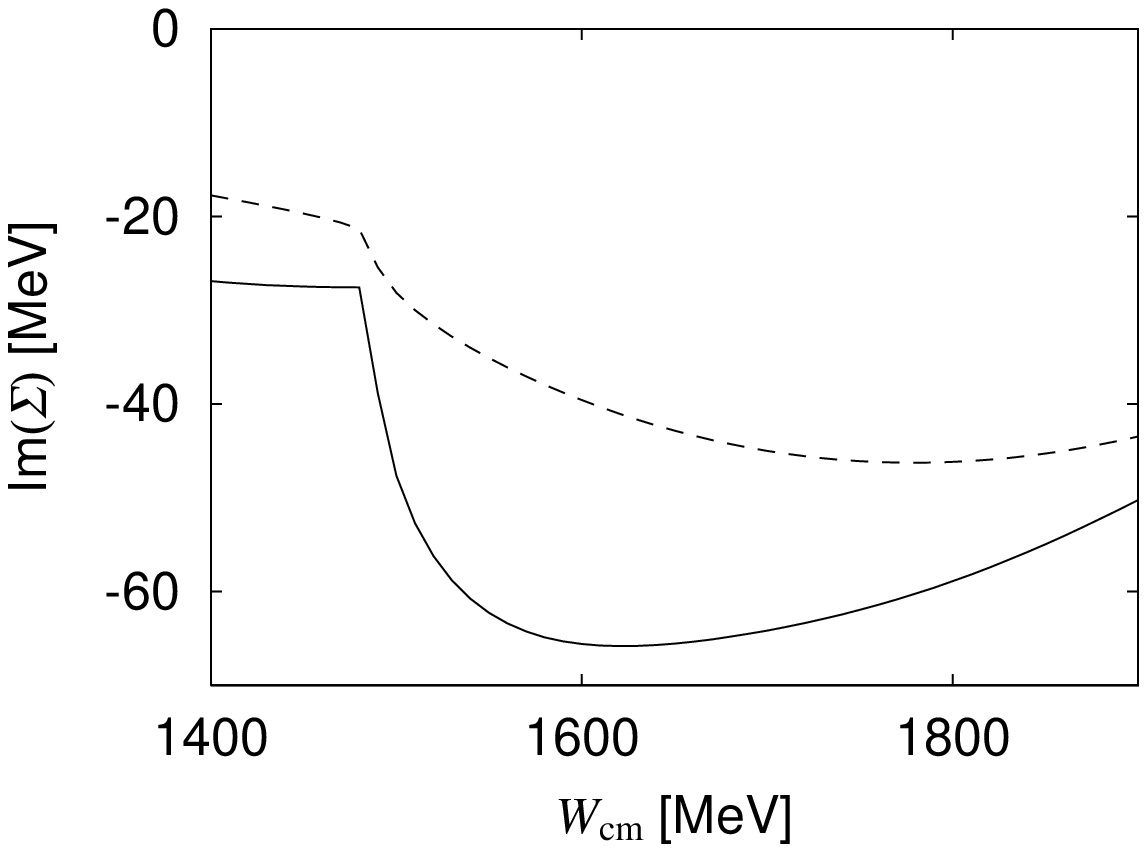,width=8cm}}
  \centerline{(a)}
  \centerline{\epsfig{file=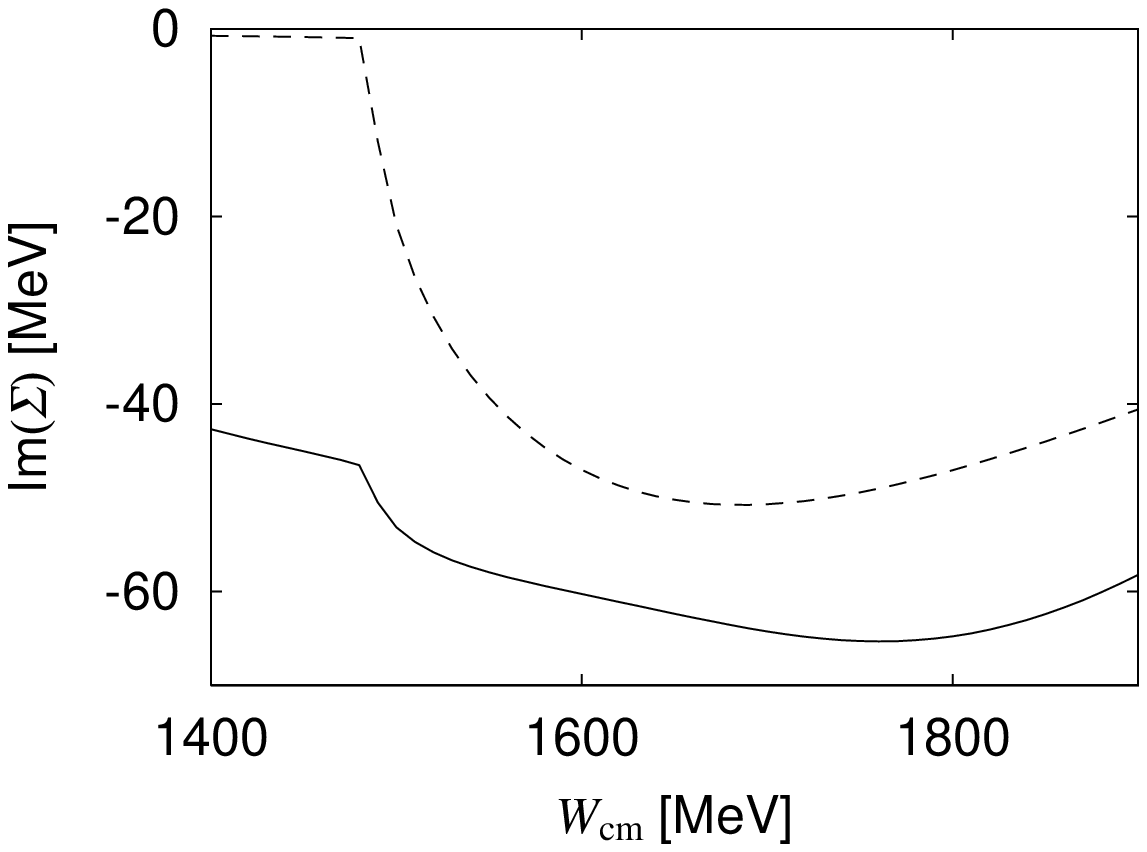,width=8cm}}
  \centerline{(b)}\vspace{5mm}
  \caption[]{Same as Fig.\ \ref{fig19a} except for the imaginary parts
   of the eigenvalues of $D(E)$.}
    \label{fig20a}
\end{figure}

\begin{figure}
  \centerline{\epsfig{file=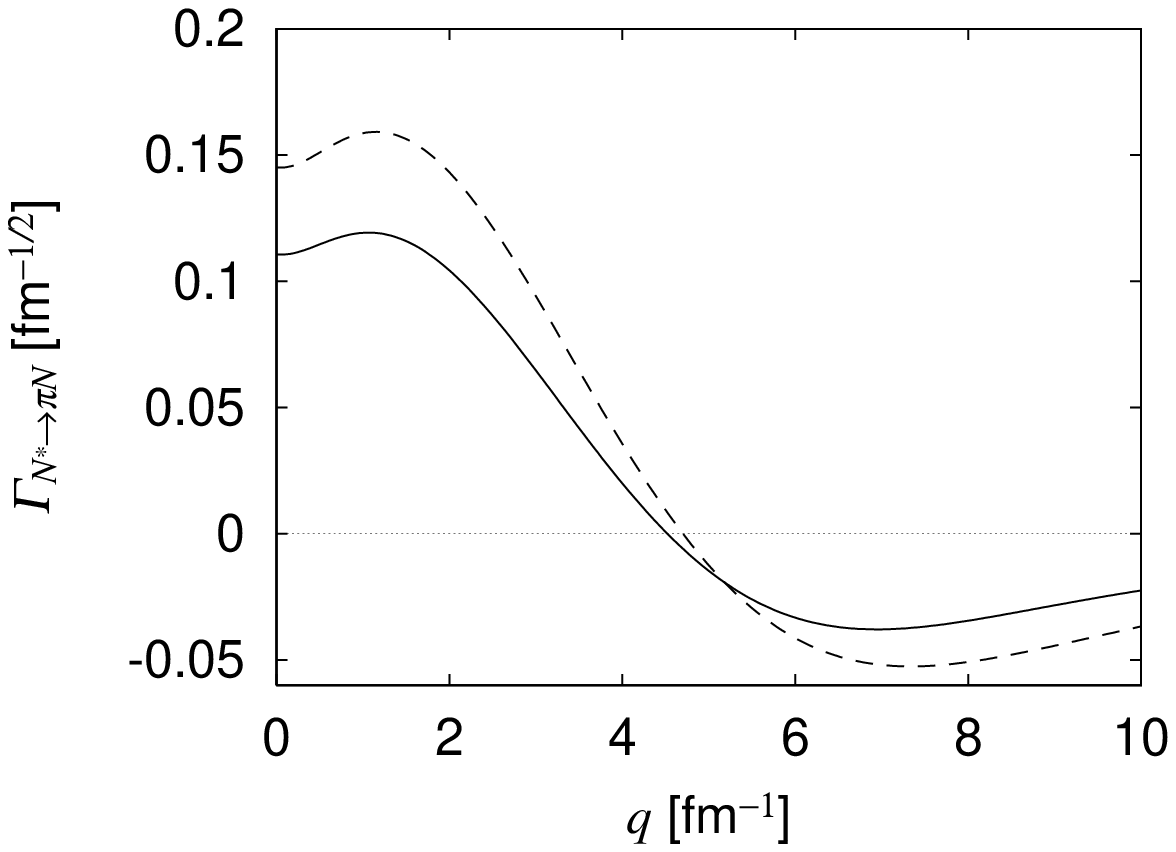,width=8cm}}
  \centerline{(a)}
  \centerline{\epsfig{file=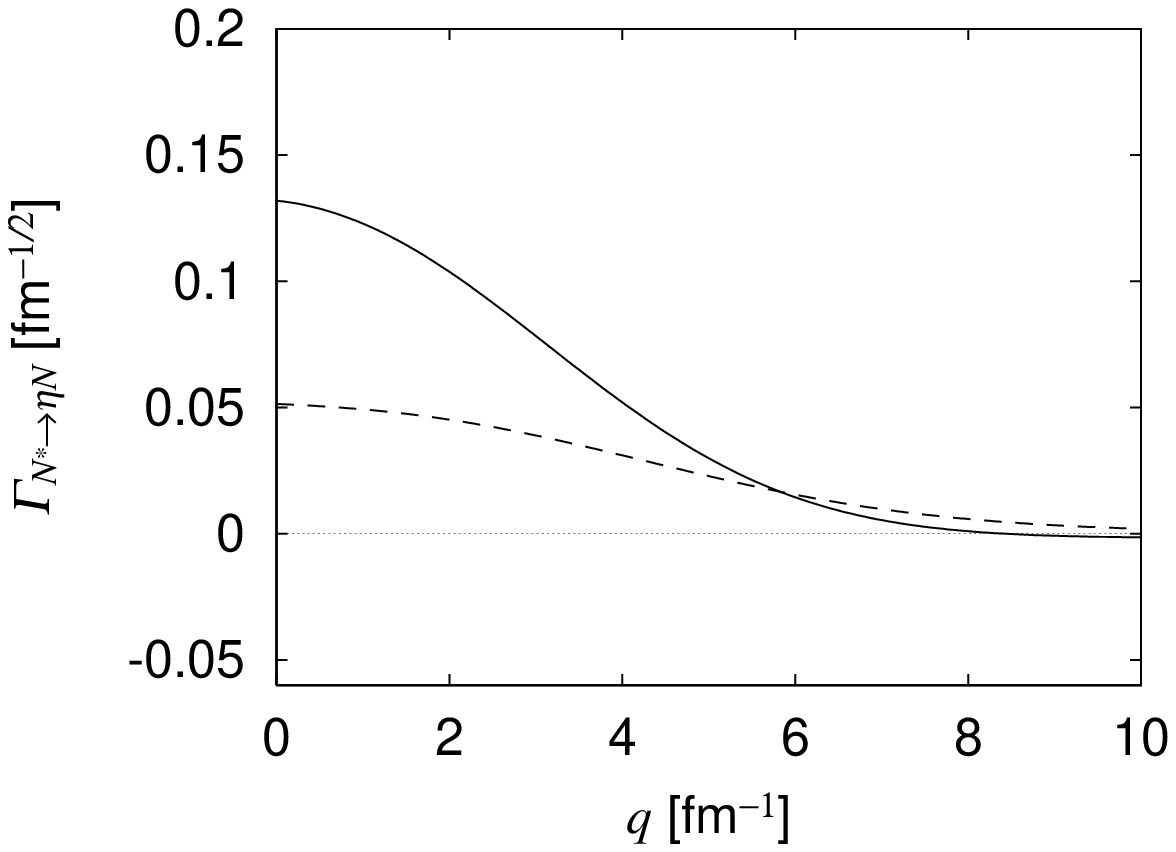,width=8cm}}
  \centerline{(b)}
  \centerline{\epsfig{file=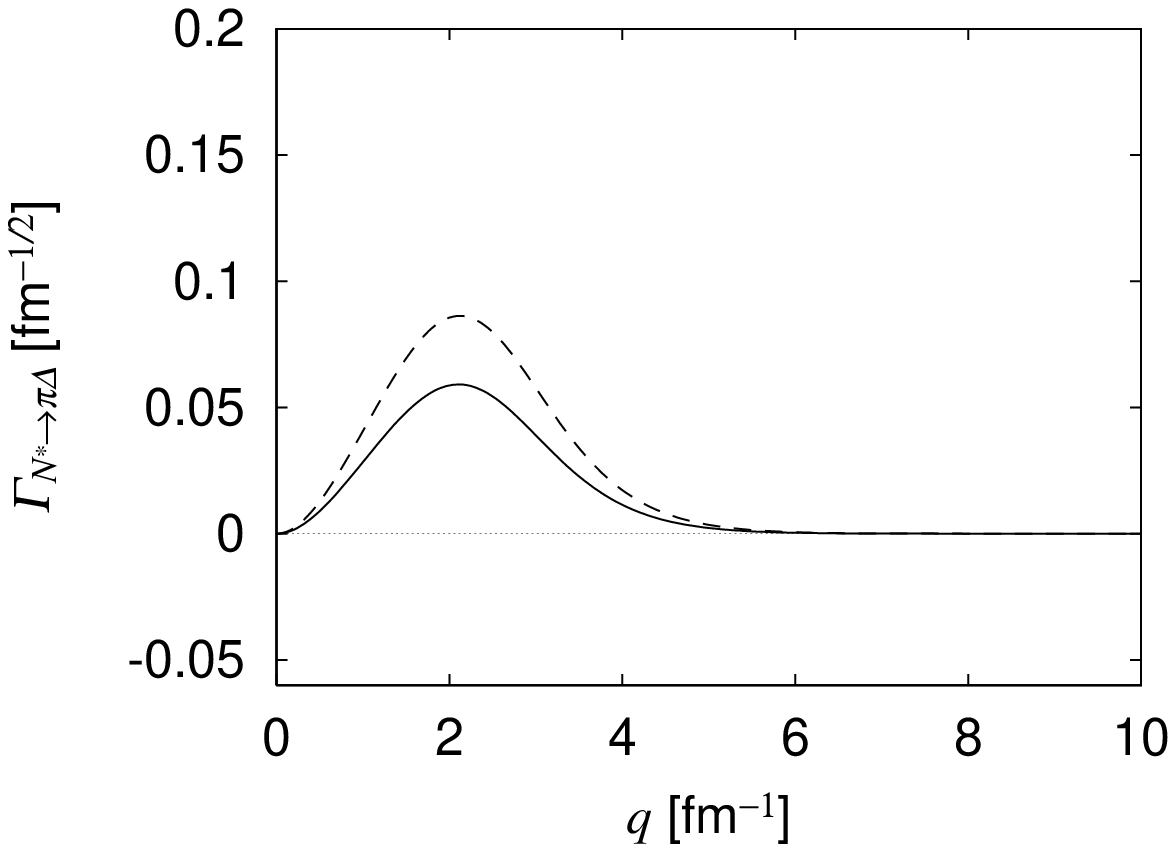,width=8cm}}
  \centerline{(c)}\vspace{5mm}
  \caption[]{$N^* \rightarrow \pi N$ (a), $N^* \rightarrow \eta N$ (b),
    and $N^* \rightarrow \pi \Delta$ (c) vertex
    functions in the phenomenological model.
    The solid and dashed curves are for $N^*_L$ and
    $N^*_H$, respectively.}
    \label{fig21a}
\end{figure}

\begin{figure}
  \centerline{\epsfig{file=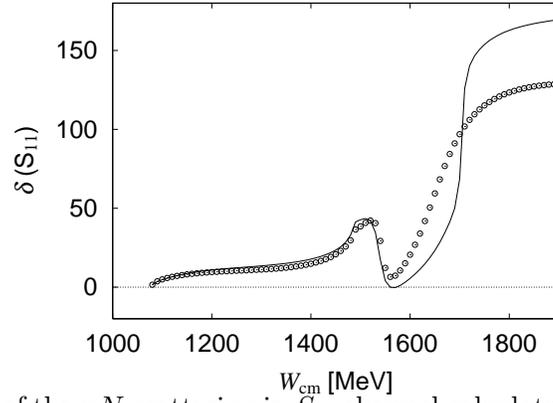,width=8cm}}
  \caption[]{Phase shifts of the $\pi N$  scattering in $S_{11}$ channel
   calculated from the phenomenological model. Open circles
   are data from VPI SP98 }
    \label{fig22}
\end{figure}

\begin{figure}
  \centerline{\epsfig{file=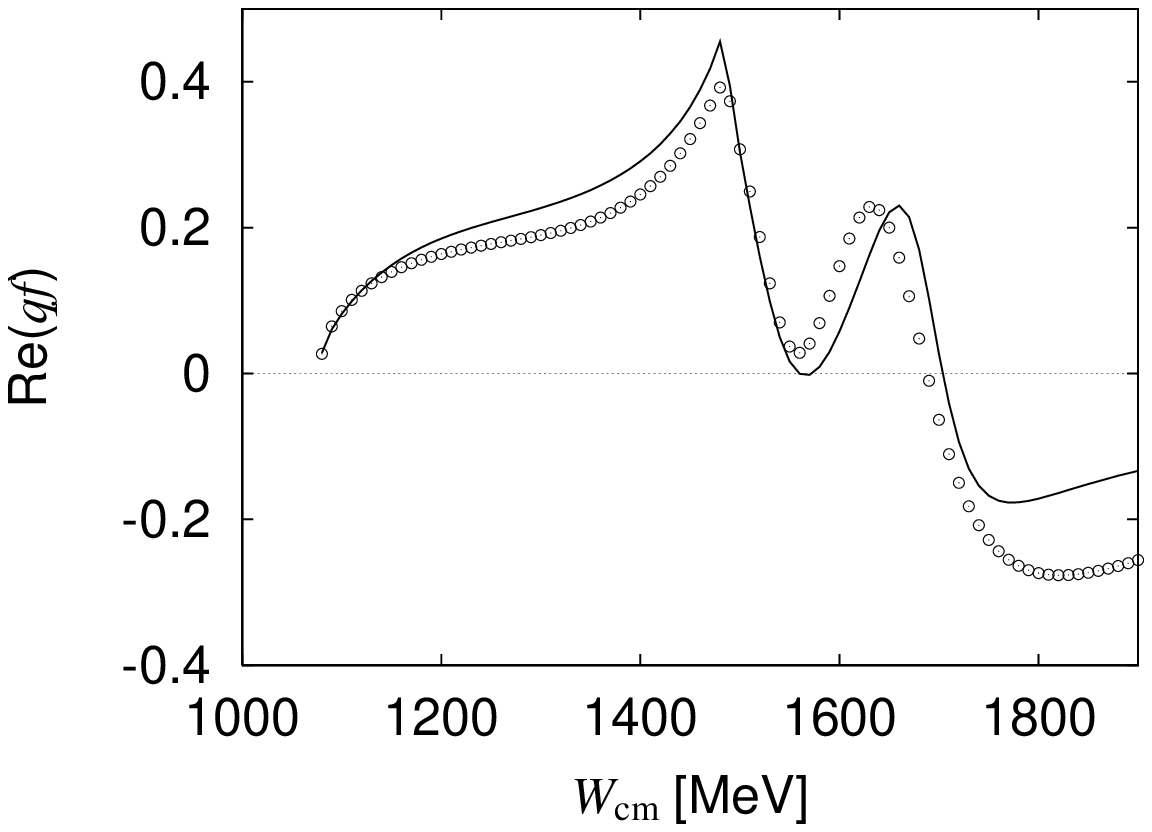,width=8cm}}
  \centerline{(a)}
  \centerline{\epsfig{file=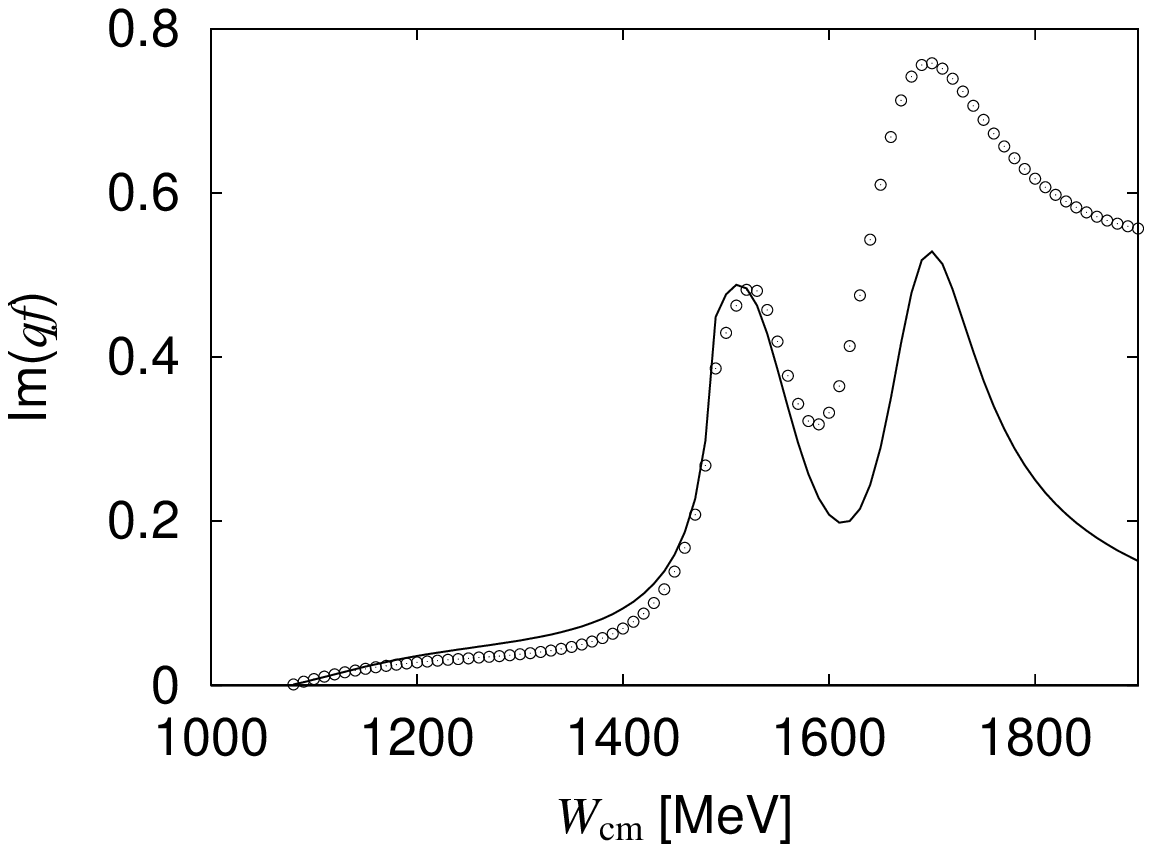,width=8cm}}
  \centerline{(b)}\vspace{5mm}
  \caption[]{Same as Fig.\ \ref{fig22} except for the
     real parts (a) and imaginary parts (b) of the $\pi N$ scattering
    amplitudes of in $S_{11}$.}
    \label{fig23}
\end{figure}

\begin{figure}
  \centerline{\epsfig{file=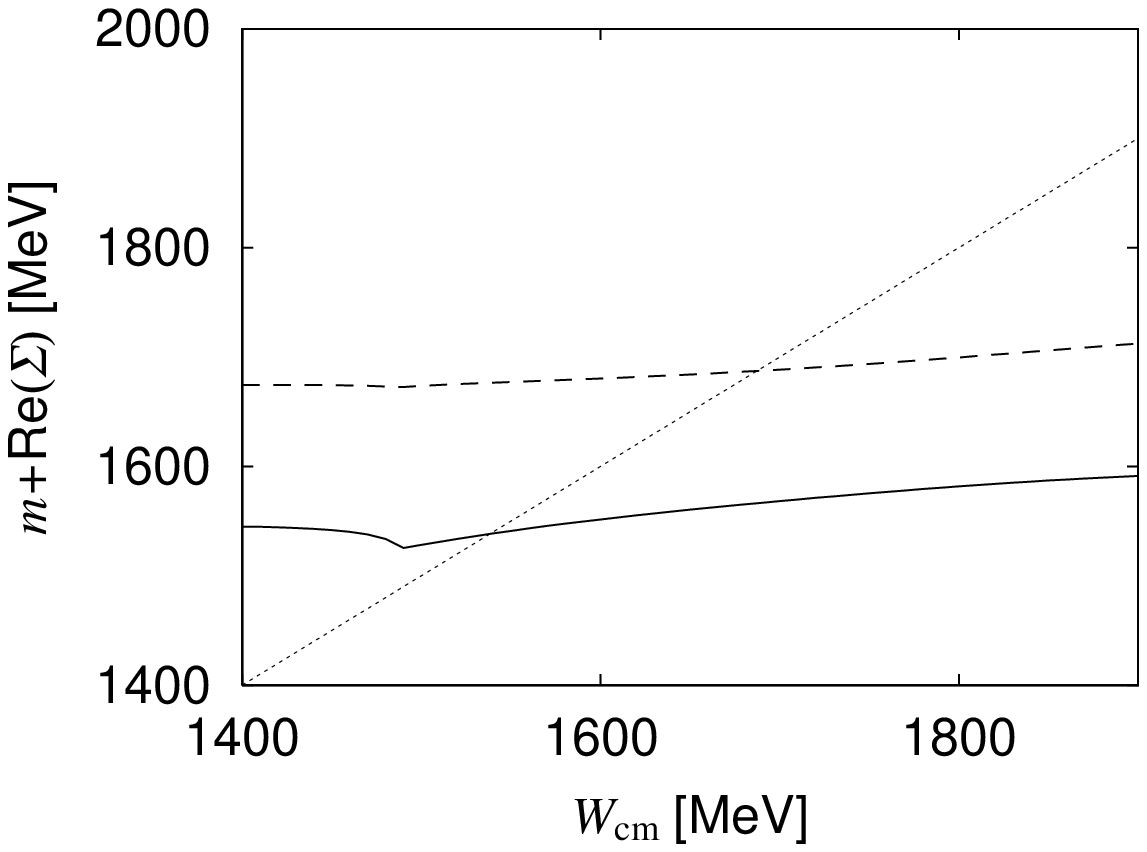,width=8cm}}
  \centerline{(a)}
  \centerline{\epsfig{file=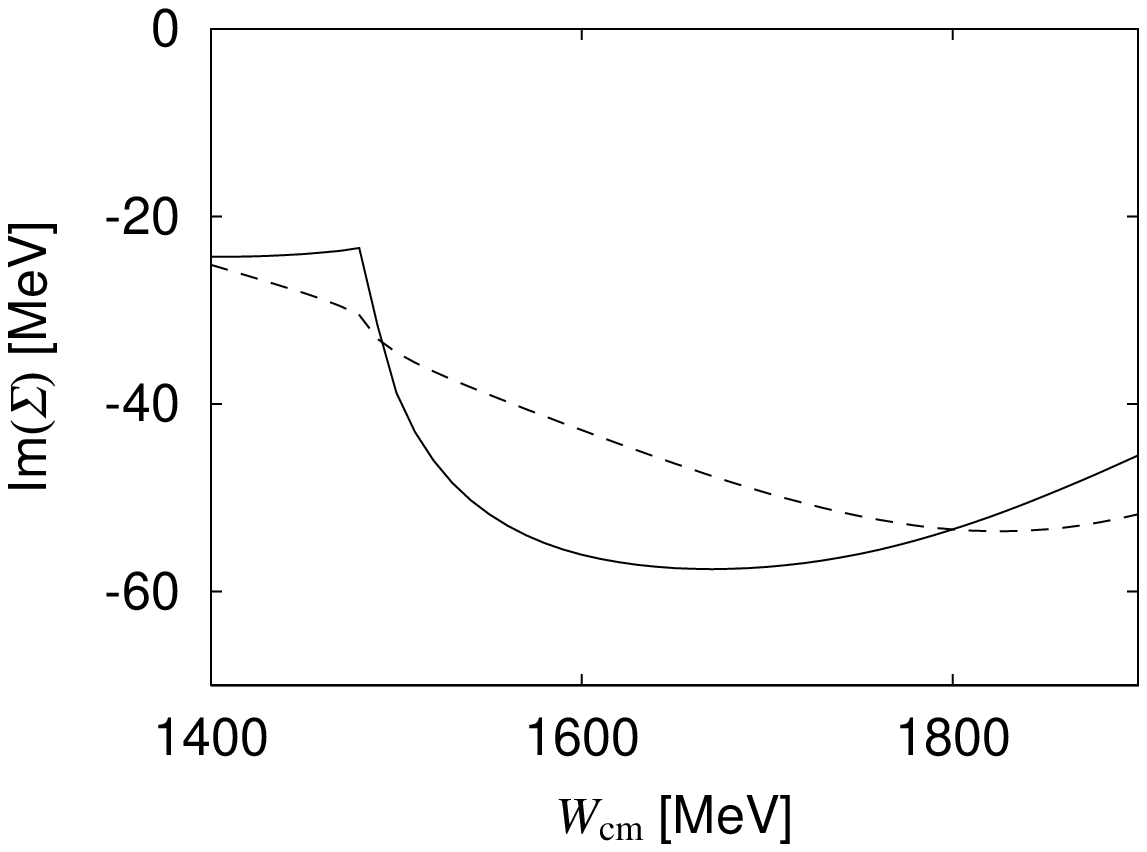,width=8cm}}
  \centerline{(b)}\vspace{5mm}
  \caption[]{Real parts (a) and imaginary parts (b) of the eigenvalues
    of $D(E)$(Eq.(63)) calculated from the phenomenological model.
  The solid and dashed curves correspond to the masses for
$N^*_L$ and $N^*_H$, respectively.}
    \label{fig24a}
\end{figure}

\begin{figure}
  \centerline{\epsfig{file=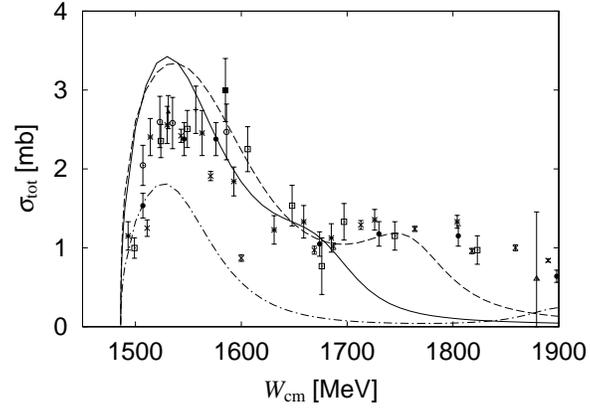,width=8cm}}\vspace{5mm}
  \caption[]{Total cross section of $\pi^- + p \rightarrow \eta + n$ reaction.
    The dashed, dot-dashed and solid curves are results of the OME,
    OGE and phenomenological models. Data are taken from
    Ref. \cite{etatot}.}
    \label{fig25}
\end{figure}

We have found that both the OGE and OME models can reproduce
the $S_{11}$ scattering amplitudes only up to about $W=$ 1500
MeV. The OGE model is better in reproducing the
rapid change in $\pi N$ phase near the $\eta$ 
production threshold, as shown in Fig.\ \ref{fig14}. However, both models
fail to describe the data at higher energies. The dynamical
origins of the difficulties are found to be due to the sensitivities
of the predicted $N^*\rightarrow \pi N, \eta N$ vertex functions
to the structure of the assumed residual quark-quark interactions. 
In particular, it is found that 
the flavor dependent ($\tau_i \cdot \tau_j$)
tensor component of the OME model does not seem to be  favored by the data.
On the other hand, the OGE model can describe the data better if
its spin-spin interaction includes a flavor dependent factor.
To illustrate this, we have shown that 
the data can be reasonably well described (Figs.\ \ref{fig22}-\ref{fig23})
by a phenomenological
model which has a spin-spin interaction from the OME model and 
a tensor interaction from the OGE model.

In conclusion, our results indicate that the residual quark-quark 
interactions within the nonrelativistic constituent quark model could be
much more complicated than the simple OGE and OME mechanisms.
In the future, we need to consider relativistic effects and the
residual quark-quark interactions due to multi-gluon and/or
multi-meson exchanges. In addition, we need to investigate the
two-body effects on the calculations of $N^*\rightarrow \pi N, \eta N,
\pi \Delta$ vertex functions. Our investigations in these directions
will be published elsewhere.

\acknowledgements
The authors would like to thank Professor H. Ohtsubo and 
Professor K. Kubodera for useful discussions.
This work is partially supported by U.S. Department of Energy, Nuclear 
Physics Division, under contract No. W-31-109-ENG-38

\newpage
\begin{table}
  \caption{Baryon states included in the u-channel non-resonant potential.}
  \label{table:bgstate}
\begin{tabular}{lllll|lllll}
$B'$ & $M'$  & $B$ & $M$   & $B_n$          
     & $B'$ & $M'$  & $B$ & $M$   & $B_n$  \\ \hline
$N$  & $\pi$ & $N$ & $\pi$ & $N,\Delta,N^*$ 
     & $\Delta$  & $\pi$  & $N$ & $\pi$ & $N,\Delta$\\
$N$  & $\eta$ & $N$ & $\pi$ & $N,N^*$    
     & $\Delta$  & $\pi$  & $N$ & $\eta$ & $\Delta$\\
$N$  & $\eta$  & $N$ & $\eta$ & $N,N^*$   
     & $\Delta$  & $\pi$  & $\Delta$ & $\pi$ & $\Delta,N^*$
\end{tabular}
\end{table}

\begin{table}
   \caption{Parameters and results of masses in the OGE model.}
  \label{table:OGEP}
  \begin{tabular}{ccccc}
    $\alpha_c$ [fm$^{-2}$] & $\alpha_s$ & $\Lambda_{\rm g}$ [MeV] &
    $M(N^*_L)$ & $M(N^*_H)$\\ \hline
    4.0 & 1.0 & 1087 & 1593.5 & 1772.4 \\ 
  \end{tabular}
\end{table}
\begin{table}
  \caption{Parameters and results of masses in the OME model.}
  \label{table:OMEP}
  \begin{tabular}{ccccc}
    $\alpha_c$ & $\Lambda_\pi$ & $\Lambda_\eta$ &
    $M(N^*_L)$ & $M(N^*_H)$ \\ \hline
    2.5 & 1139 & 1000 & 1565.0 & 1885.8 
  \end{tabular}
\end{table}

\begin{table}
  \caption{Parameters of Fermi type form factors.}
  \label{table:Reac}
  \begin{tabular}{c|l|rr|rr|rr}
    &  & \multicolumn{2}{c|}{$\pi N$}  &
    \multicolumn{2}{c|}{$\eta N$} &
    \multicolumn{2}{c}{$\pi \Delta$} \\ \hline
    & $X_{\eta qq}$ & $k_0$ & $\Delta_k$ & $k_0$ &
    $\Delta_k$ & $k_0$ & $\Delta_k$\\ \hline
    OGE & 0.67 & 5 & 5 & 5 & 6.5 & 2.40 & 0.60 \\
    OME & 1.00 & 5 & 5 & 5 & 6.5 & 2.43 & 0.55 \\
    Phenom. & 0.48 & 10 & 7 & 10 & 6 & 2.45 & 0.60 
  \end{tabular}
\end{table}
\begin{table}
  \caption{Coefficients of 1$\hbar \omega$ configurations.}
  \label{table:Coef}
  \begin{tabular}{c|rr|rr|rr}
    & \multicolumn{2}{c|}{OGE}    & \multicolumn{2}{c|}{OME} &
      \multicolumn{2}{c}{Phenom.}  \\
    & $N^*_L$  & $N^*_H$   & $N^*_L$   & $N^*_H$  &
      $N^*_L$  & $N^*_H$ \\ \hline
    $S=1/2$ & $0.8075$ & $-0.1992$ & $-0.7463$ & $0.2093$ &
              $0.7081$ & $-0.3074$\\
    $S=3/2$ & $0.2325$ & $0.7552$  &  $0.2948$ & $0.6034$ &
              $0.3846$  & $0.6623$ 
  \end{tabular}
\end{table}

\begin{table}
  \caption{Parameters and results of masses in the phenomenological model.}
  \label{table:PHEN}
  \begin{tabular}{cccccc}
    $\alpha_c$ & $\alpha_{\sigma \tau}$ &
    $\alpha_T$ & $\Lambda_{\rm ph}$ & $M(N^*_L)$ & $M(N^*_H)$\\ \hline
    2.5 & 0.427 & 1.60 & 1200 & 1580.6 & 1719.5
  \end{tabular}
\end{table}

                                %

\newpage
\newpage
\newpage
\newpage
\newpage
\newpage
\newpage
\newpage
\newpage
\newpage
\newpage
\newpage
\newpage
\newpage

\newpage                                %

\end{document}